\documentclass[twocolumn]{aastex631}

\usepackage{url}
\usepackage{threeparttable}
\usepackage{ulem}
\usepackage{natbib}
\usepackage{appendix}
\usepackage{amsmath}
\usepackage{amssymb}
\usepackage{multirow}
\usepackage{float}
\usepackage{physics}

\def\ion#1#2{#1$\;${\sc\@roman{#2}}\relax}
\def\lesssim{\mathrel{\hbox{\rlap{\hbox{\lower4pt\hbox{$\sim$}}}\hbox{$<$}}}}
\def\gtrsim{\mathrel{\hbox{\rlap{\hbox{\lower4pt\hbox{$\sim$}}}\hbox{$>$}}}}

\newcommand{\dvLya}{\Delta v_{\mathrm{Ly}\alpha}}
\newcommand{\xHI}{x_{\rm H\,\textsc{i}}}
\newcommand{\Lya}{\mathrm{Ly}\alpha}
\newcommand{\EWLya}{\mathrm{EW}_{0,\Lya}}
\newcommand{\meanz}{\langle z\rangle}
\newcommand{\XLya}{X_\mathrm{Ly\alpha}}
\newcommand{\arcsecond}{\ensuremath{''}}

\shorttitle{Ly$\alpha$ Emission at $z=7-13$ and Cosmic Reionization}
\shortauthors{Nakane et al.}

\begin{document}

\title{
Ly$\alpha$ Emission at $z=7-13$: Clear Ly$\alpha$ Equivalent Width Evolution\\ Indicating the Late Cosmic Reionization History


}

\author[0009-0000-1999-5472]{Minami Nakane}
\affiliation{Institute for Cosmic Ray Research, The University of Tokyo, 5-1-5 Kashiwanoha, Kashiwa, Chiba 277-8582, Japan}
\affiliation{Department of Physics, Graduate School of Science, The University of Tokyo, 7-3-1 Hongo, Bunkyo, Tokyo 113-0033, Japan}

\author[0000-0002-1049-6658]{Masami Ouchi}
\affiliation{National Astronomical Observatory of Japan, 2-21-1 Osawa, Mitaka, Tokyo 181-8588, Japan}
\affiliation{Institute for Cosmic Ray Research, The University of Tokyo, 5-1-5 Kashiwanoha, Kashiwa, Chiba 277-8582, Japan}
\affiliation{Department of Astronomical Science, SOKENDAI (The Graduate University for Advanced Studies), Osawa 2-21-1, Mitaka, Tokyo, 181-8588, Japan}
\affiliation{Kavli Institute for the Physics and Mathematics of the Universe (WPI), University of Tokyo, Kashiwa, Chiba 277-8583, Japan}

\author[0000-0003-2965-5070]{Kimihiko Nakajima}
\affiliation{National Astronomical Observatory of Japan, 2-21-1 Osawa, Mitaka, Tokyo 181-8588, Japan}

\author[0000-0002-6047-430X]{Yuichi Harikane} 
\affiliation{Institute for Cosmic Ray Research, The University of Tokyo, 5-1-5 Kashiwanoha, Kashiwa, Chiba 277-8582, Japan}

\author[0000-0001-9011-7605]{Yoshiaki Ono}
\affiliation{Institute for Cosmic Ray Research, The University of Tokyo, 5-1-5 Kashiwanoha, Kashiwa, Chiba 277-8582, Japan}

\author[0009-0008-0167-5129]{Hiroya Umeda}
\affiliation{Institute for Cosmic Ray Research, The University of Tokyo, 5-1-5 Kashiwanoha, Kashiwa, Chiba 277-8582, Japan}
\affiliation{Department of Physics, Graduate School of Science, The University of Tokyo, 7-3-1 Hongo, Bunkyo, Tokyo 113-0033, Japan}

\author[0000-0001-7730-8634]{Yuki Isobe}
\affiliation{Institute for Cosmic Ray Research, The University of Tokyo, 5-1-5 Kashiwanoha, Kashiwa, Chiba 277-8582, Japan}
\affiliation{Department of Physics, Graduate School of Science, The University of Tokyo, 7-3-1 Hongo, Bunkyo, Tokyo 113-0033, Japan}

\author[0000-0003-3817-8739]{Yechi Zhang}
\affiliation{Institute for Cosmic Ray Research, The University of Tokyo, 5-1-5 Kashiwanoha, Kashiwa, Chiba 277-8582, Japan}

\author[0000-0002-5768-8235]{Yi Xu}
\affiliation{Institute for Cosmic Ray Research, The University of Tokyo, 5-1-5 Kashiwanoha, Kashiwa, Chiba 277-8582, Japan}
\affiliation{Department of Astronomy, Graduate School of Science, The University of Tokyo, 7-3-1 Hongo, Bunkyo, Tokyo 113-0033, Japan}



\begin{abstract}
We present the evolution of Ly$\alpha$ emission derived from 53 galaxies at $z=6.6-13.2$ that are identified by multiple JWST/NIRSpec spectroscopy programs of ERS, GO, DDT, and GTO. These galaxies fall on the star-formation main sequence and are the typical star-forming galaxies with UV magnitudes of $-22.5\leq M_\mathrm{UV}\leq-17.0$.
We find that 15 out of 53 galaxies show Ly$\alpha$ emission at the $>3\sigma$ levels, and obtain Ly$\alpha$ equivalent width (EW) measurements and stringent $3\sigma$ upper limits for the 15 and 38 galaxies, respectively. Confirming that Ly$\alpha$ velocity offsets and line widths of our galaxies are comparable with those of low-redshift Ly$\alpha$ emitters, we investigate the redshift evolution of the Ly$\alpha$ EW.
We find that Ly$\alpha$ EWs statistically decrease towards high redshifts on the Ly$\alpha$ EW vs. $M_{\rm UV}$ plane for various probability distributions of the uncertainties.
We then evaluate neutral hydrogen fractions $\xHI$ with the Ly$\alpha$ EW redshift evolution and the cosmic reionization simulation results on the basis of a Bayesian inference framework, and obtain $\xHI<0.79$, $=0.62^{+0.15}_{-0.36}$, and $0.93^{+0.04}_{-0.07}$ at $z\sim7$, $8$, and $9-13$, respectively. These moderately large $\xHI$ values are consistent with the Planck CMB optical depth measurement and previous $\xHI$ constraints from galaxy and QSO Ly$\alpha$ damping wing absorptions, and strongly indicate a late reionization history. Such a late reionization history suggests that major sources of reionization would emerge late and be hosted by moderately massive halos in contrast with the widely-accepted picture of abundant low-mass objects for the sources of reionization.


\end{abstract}



\section{Introduction} \label{sec:intro}

Cosmic reionization is a crucial phase transition in the early universe. During the epoch of reionization, the neutral hydrogen in the inter-galactic medium (IGM) was ionized by the UV photons emitted from the first sources. \citet{Fan2006} show that reionization mostly completed by $z\sim6$ from the measurements of \citet{Gunn&Peterson1965} troughs in the spectra of quasars (QSOs) while \citet{Kulkarni2019} suggest that a broad scatter in the neutral hydrogen fraction $\xHI$ exists at redshift as low as $z\lesssim5.5$.  Thomson scattering optical depth to the cosmic microwave background (CMB) indicates that significant reionization occurred at $z\sim7-8$ \citep{Planck2020}. 

Reionziation history is investigated by deriving the redshift evolution of $\xHI$ in the IGM. $\Lya$ emission from star-froming galaxies is a good probe of $\xHI$ because $\Lya$ emission line is strongly attenuated by the neutral hydrogen via the $\Lya$ damping wing \citep{Ouchi2020}. While Lyman-alpha emitters (LAEs) have been commonly found at $z\simeq6$ (e.g., \citealt{Stark2011,DeBarros2017}), LAEs at higher redshift $z\gtrsim7$ are rare (e.g., \citealt{Stark2010,Treu2013,Hoag2019,Jung2020}). Although reionization history is constrained by many studies of LAE clustering analysis (e.g., \citealt{Ouchi2018,U23_Thesis}), $\Lya$ luminosity function (e.g., \citealt{Ouchi2010,Konno2014,Hu2019,Wold2022}), and $\Lya$ EW distribution (e.g., \citealt{Mason2018,Hoag2019,Bolan2022}), the uncertainties of reionization history are still large at $z\sim6-8$. At higher redshift $z\gtrsim8$, almost no observational constraints exist before the arrival of James Webb Space Telescope (JWST; \citealt{Gardner2023,Rigby2023,Rieke2023,Jakobsen2022,Ferruit2022,Boker2023}). JWST allows a deep survey of faint galaxies during the epoch of reionization. In fact, many LAEs are spectroscopically identified at high-redshift $7\lesssim z\lesssim11$ (e.g., \citealt{Tang2023,Bunker2023a,Jung2023,Jones2023,Saxena2023,Chen2024}). In addition, recent studies of the JWST observations present the constraints on reionization history at $z\gtrsim8$ \citep{Curtis-Lake2023,Hsiao2023,Umeda2023} although the constraints are not strong. The open questions of reionization are thus when reionization started and at what pace reionization proceeded.

$\Lya$ EW defined by a $\Lya$ flux relative to a UV continuum flux is a key observable property that reflects the intensity of the $\Lya$ emission. The fraction of LAEs ($\mathrm{EW}>25$\AA) among Lyman-break galaxies (LBGs), the $\Lya$ fraction, is measured as a way to constraint on the $\Lya$ optical depth (e.g., \citealt{Stark2010,Ono2012,Schenker2012,Treu2012,Pentericci2018,Jones2023}). The $\Lya$ fraction dramatically decrease from $z\sim6$ to $7$, indicating the rapid increase in $\Lya$ optical depth at $z\gtrsim6$. \citet{Treu2012,Treu2013} suggest that the full $\Lya$ EW distribution including galaxies with low $\Lya$ EW and no $\Lya$ detections contains more information about the $\Lya$ optical depth rather than the $\Lya$ fraction, which evaluate the fraction of galaxies detected at some fixed threshold of $\Lya$ EW. The $\xHI$ values in the IGM are not directly constrained by the $\Lya$ optical depth because $\Lya$ photons are absorbed by the neutral hydrogen not only in the IGM, but also in the inter-stellar medium (ISM) and circum-galactic medium (CGM). The effects of the ISM and CGM on $\Lya$ transmission are studied by some models or simulations (e.g., \citealt{Dijkstra2011,Mesinger2015}). \citet{Mason2018} introduce a flexible Bayesian framework based on that of \citet{Treu2012,Treu2013} and the simulations by \citet{Mesinger2015} to infer $\xHI$ from LBGs with detections and no detections of $\Lya$ emission.

Reionization source is another important feature of reionization. Recent observational studies report a steep faint end slope $\alpha$ of UV luminosity function \citep{Bouwens2017,Livermore2017,Atek2018,Ishigaki2018,Finkelstein2019}, indicating that faint sources make dominant contributions to UV luminosity correlated with ionizing photon production. Although the escape fraction of ionizing photons $f_\mathrm{esc}$ is also crucial to understanding ionizing sources, it is impossible to directly measure $f_\mathrm{esc}$ at $z>4$ due to the high optical depth of ionizing photons (e.g., \citealt{Vanzella2018}). In addition, recent works of simulations and semi-analytical models show conflicting results for $f_\mathrm{esc}$ correlated with halo mass (e.g., \citealt{Gnedin2008,Wise2009,Sharma2016}) and anti-correlated with halo mass (e.g., \citealt{Paardekooper2015,Faisst2016,Kimm2017}). The $f_\mathrm{esc}$ value during the epoch of reionization thus remains uncertain. However, given that low-mass (massive) galaxies contribute to early (late) reionization (e.g., \citealt{Faisst2016,Dayal2024}), we may be able to constrain on the mass dependence of reionization source from reionization history.

There are three scenarios of the reionization history suggested by \citet{Naidu2020}, \citet{Ishigaki2018}, and \citet{Finkelstein2019}, referred to as the late scenario, medium-late scenario, and early scenario, respectively. The three scenarios differ in the escape fraction of ionizing photons $f_\mathrm{esc}$ and the faint-end slope $\alpha$ of UV luminosity function. We refer to Model II described in \citet{Naidu2020} as the late scenario because Model II self-consistently explain observational results of the redshift evolution of $f_\mathrm{esc}$ (e.g., \citealt{Siana2010,Rutkowski2016,Marchi2017,Steidel2018,Fletcher2019}). In the late scenario, the faint-end slope should be shallow ($\alpha>-2$) with the assumption that $f_\mathrm{esc}$ depends on the star-formation rate surface density $\Sigma_\mathrm{SFR}$. Bright and massive galaxies ($M_\mathrm{UV}<-18$ and stellar mass $M_*>10^8M_\odot$) are major contributors to reionization in this scenario. In the medium-late scenario, \citet{Ishigaki2018} find the steep faint-end slope ($\alpha\lesssim-2$) from observations and the constant escape fraction ($f_\mathrm{esc}\sim0.17$) with which the scenario can mostly explain the observational measurements. Galaxies with various luminosities contribute to reionization in this scenario. In the early scenario, the faint-end slope should be steep ($\alpha\lesssim-2$) with the assumption that $f_\mathrm{esc}$ anti-correlate with halo mass $M_\mathrm{h}$. Faint and low-mass galaxies ($M_\mathrm{UV}>-15$ and $M_*\lesssim10^6M_\odot$) are major contributors to reionization in this scenario. 

In this study, we present the evolution of $\Lya$ emission with 53 galaxies at $z\gtrsim7$ whose redshifts are spectroscopically confirmed by JWST/NIRSpec. Combining these high-redshift galaxies and the full distributions of $\Lya$ EW with detections and upper limits, we infer $\xHI$ at $z\sim7-13$. The $\xHI$ values provide insights into reionization history and sources. This paper is constructed as follows. Section \ref{sec:datasam} describe the JWST/NIRSpec spectroscopic data and our sample of galaxies. We conduct spectral fittings to our sample galaxies and investigate the $\Lya$ properties obtained from the best-fit parameters in Section \ref{sec:measure}. In Section \ref{sec:xHI}, we explain our method to infer $\xHI$ and present the estimated $\xHI$ values. In Section \ref{sec:diss}, we discuss reionization history and sources. Section \ref{sec:sum} summarizes our findings.  Throughout this paper, we use a standard $\Lambda$CDM cosmology with $\Omega_\Lambda=0.7$, $\Omega_m=0.3$, and $H_0=70$ km $\mathrm{s}^{-1}$ $\mathrm{Mpc}^{-1}$. All magnitudes are in the AB system \citep{Oke&Gunn1983}.

\section{Data and Sample} \label{sec:datasam}

The spectroscopic data sets used in this study were obtained in multiple programs of the public observations; the Early Release Science (ERS) observations of GLASS (ERS-1324, PI: T. Treu; \citealt{Treu2022}) and the Cosmic Evolution Early Release Science (CEERS; ERS-1345, PI: S. Finkelstein; \citealt{Finkelstein2023}, \citealt{Haro2023a}), General Observer (GO) observations (GO-1433, PI: D. Coe; \citealt{Hsiao2023}), the Director’s Discretionary Time (DDT) observations (DDT-2750, PI: P. Arrabal Haro; \citealt{Haro2023b}), and the Guarnteed Time Observations (GTO) of the JWST Advanced Deep Extragalactic Survey (JADES; GTO-1210, PI: N. L\"{u}tzgendorf; \citealt{Bunker2023b}, \citealt{Eisenstein2023}). The GLASS data were taken at one pointing position with high resolution ($R\sim2700$) filter-grating pairs of F100LP-G140H, F170LP-G235H, and F290LP-G395H covering the wavelength ranges of $1.0-1.6$, $1.7-3.1$, and $2.9-5.1\mu$m, respectively. The total exposure time of the GLASS data is 4.9 hours for each filter-grating pair. The CEERS data were taken at 8 and 6 pointing positions with the prism ($R\sim100$) and medium-resolution grating ($R\sim1000$), respectively. The prism covered the wavelength of $0.6-5.3$ $\mu$m. The filter-grating pairs of medium-resolution grating observations were F100LP-G140M, F170LP-G235M, and F290LP-G395M whose wavelength coverages were $1.0-1.6$, $1.7-3.1$ and $2.9-5.1$ $\mu$m, respectively. The total exposure time of the CEERS data is 0.86 hours for each filter-grating pair per pointing. The GO-1433 and DDT-2750 observations, each at a single pointing, were conducted with the prism, and the total exposure times are 3.7 and 5.1 hours, respectively. The JADES data were obtained at 3 pointing positions with the prism and 4 filter-grating pairs of F070LP-G140M ($0.7-1.3$$\mu$m), F170LP-G235M, F290LP-G395M, and F290LP-G395H. The total exposure times of the JADES observations are 9.3 hours for the prism and 2.3 hours for each filter-grating pair per pointing. \citet{Nakajima2023} and \citet{Harikane2024} have reduced the GLASS, CEERS, GO-1433, and DDT-2750 data with the JWST pipeline version 1.8.5 with the Calibration Reference Data System (CRDS) context file of \texttt{jwst\_1028.pmap} or \texttt{jwst\_1027.pmap} with additional processes improving the flux calibration, noise estimate, and the composition, and have measured systemic redshifts $z_\mathrm{sys}$ with optical emission lines (H$\beta$, [{\sc Oiii}]$\lambda$4959, and [{\sc Oiii}]$\lambda$5007). See \citet{Nakajima2023} and \citet{Harikane2024} for further details of the data reduction and $z_\mathrm{sys}$ determination. The combination of the data of \citet{Nakajima2023} and \citet{Harikane2024} from the public observational programs provides a sample of 196 galaxies at $4.0\leq z_\mathrm{sys}\leq11.4$. We then select 35 galaxies with $z_\mathrm{sys}\geq7$ from the sample that are referred to as the public-data sample. The JADES data are publicly available \footnote{\url{https://archive.stsci.edu/hlsp/jades}}, and already reduced with the pipeline developed by the ESA NIRSpec Science Operations Team and the NIRSpec GTO Team. The JADES data consists of 178 galaxies at $0.7\leq z_\mathrm{sys}\leq13.2$ including 75 galaxies whose redshifts are not determined. We select 18 galaxies from the JADES data with $z_\mathrm{sys}\geq6.5$ identified by \citet{Bunker2023b} with emission lines (e.g., H$\beta$, [{\sc Oiii}]$\lambda$4959, and [{\sc Oiii}]$\lambda$5007) or spectral breaks. To these 18 galaxies, we add GN-z11 at $z_\mathrm{sys}=10.6$ that is also observed as part of the JADES program (GTO-1181, PI: D. Eisenstein; \citealt{Eisenstein2023}). We refer a total of these 19 galaxies as the JADES-data sample. Note that we use the values of the rest-frame $\Lya$ equivalent width, velocity offset, and line width presented in the literature \citep{Bunker2023a}. Combining the JADES-data sample with the public-data sample, we obtain a total of 54 galaxies whose $z_\mathrm{sys}$ fall in the range of $6.6\leq z_\mathrm{sys}\leq13.2$. Here, 6, 26, 1, 2, and 19 out of all 54 galaxies are found in the GLASS, CEERS, GO-1433, DDT-2750, and JADES data, respectively. We note that these 54 galaxies in the public-data and JADES-data samples are observed with the gratings and/or prism. The public-data sample (35 galaxies) consists of 17, 26, and 8 galaxies that have the grating, prism, and grating+prism spectra, respectively. The JADES-data sample (19 galaxies) has both the grating and prism spectra, while the grating spectra at the $\Lya$ wavelengths are missing in 3 galaxies. There are a total of 33 ($=17+19-3$) galaxies whose $\Lya$ emission wavelengths are observed with the gratings. The rest of the objects, 21 ($=54-33$) galaxies, have only the prism spectra, where the measurements of the $\Lya$ velocity offset and line width are not obtained due to the poor spectral resolution.

\begin{figure*}[t]
    \centering
    \includegraphics[scale=0.42]{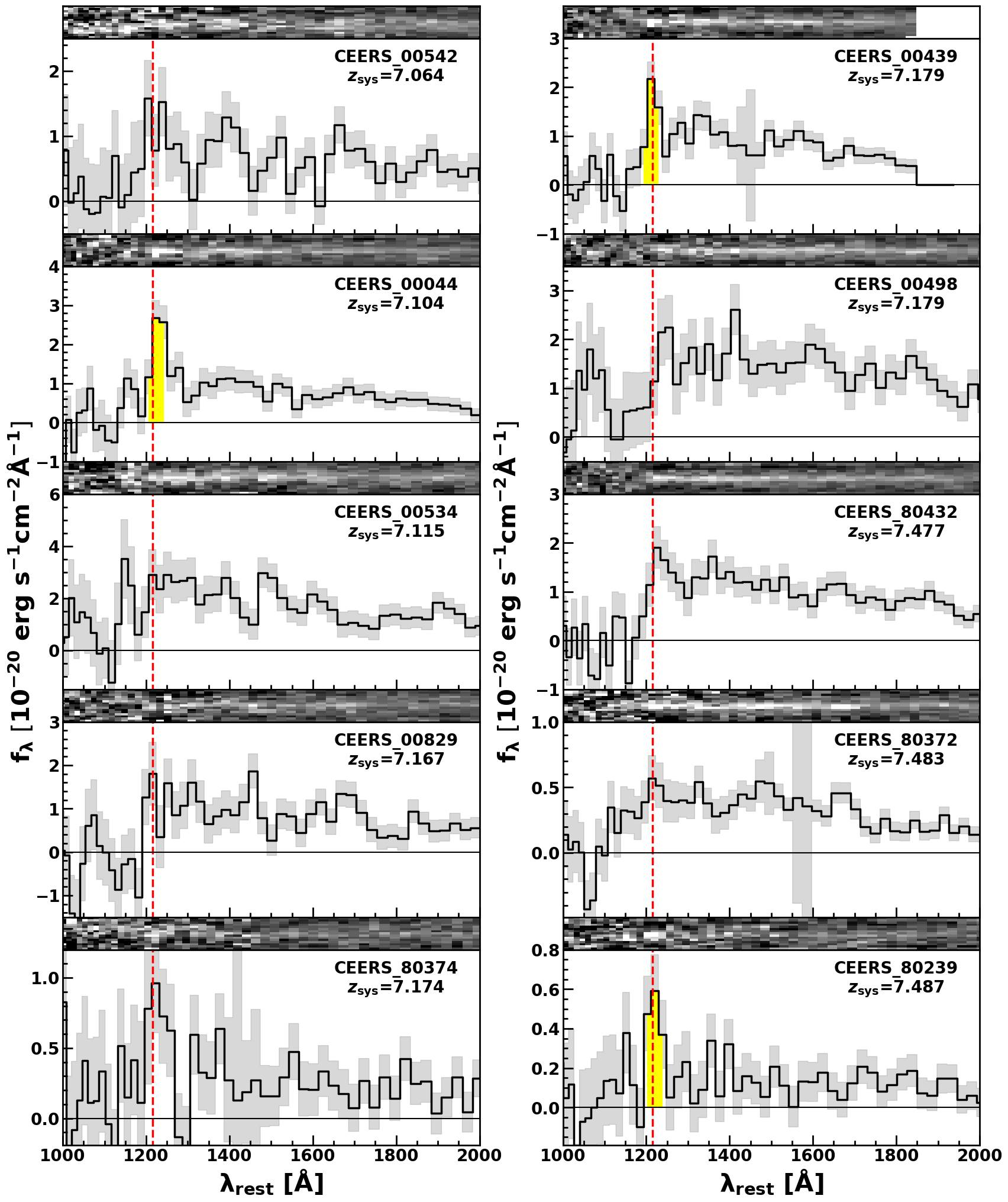}
    \caption{Spectra of the public-data sample. 
    Each panel shows the two-dimensional spectrum (top) and the one-dimensional spectrum (bottom). The black solid lines and shaded regions represent the observed spectra and associated $1\sigma$ errors, respectively. The vertical red dashed lines represent the rest-frame $\Lya$ wavelengths of $1215.67$ \AA\ at the systemtic redshifts. The yellow regions indicate the detected $\Lya$ lines whose signal to noise ratio is larger than $3\sigma$ (Section \ref{subsec:fitting}).}
    \label{fig:public-data_sample}
\end{figure*}

\setcounter{figure}{0}
\begin{figure*}[t]
    \centering
    \includegraphics[scale=0.42]{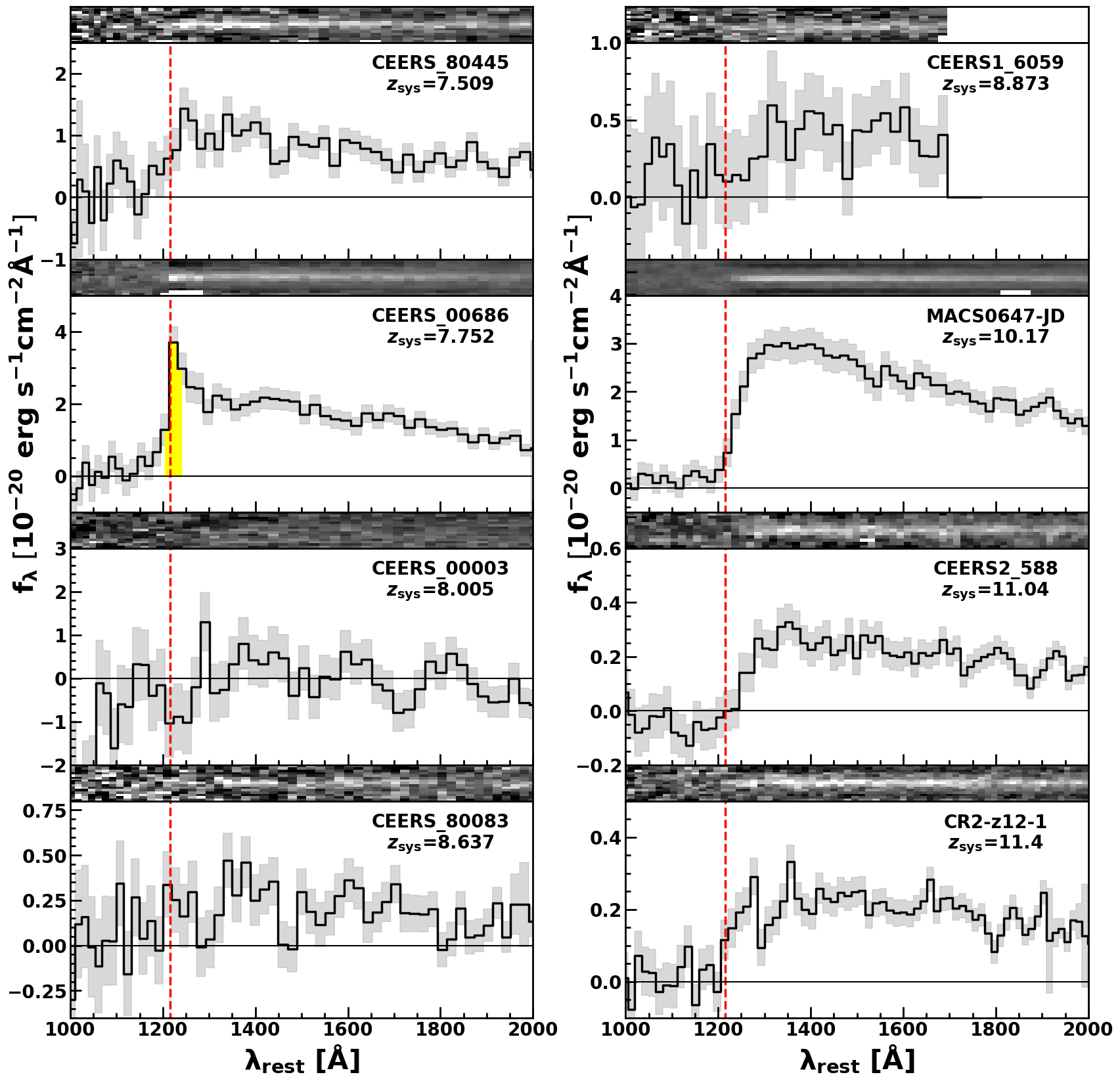}
    \caption{Continued.}
\end{figure*}

\setcounter{figure}{0}
\begin{figure*}[t]
    \centering
    \includegraphics[scale=0.42]{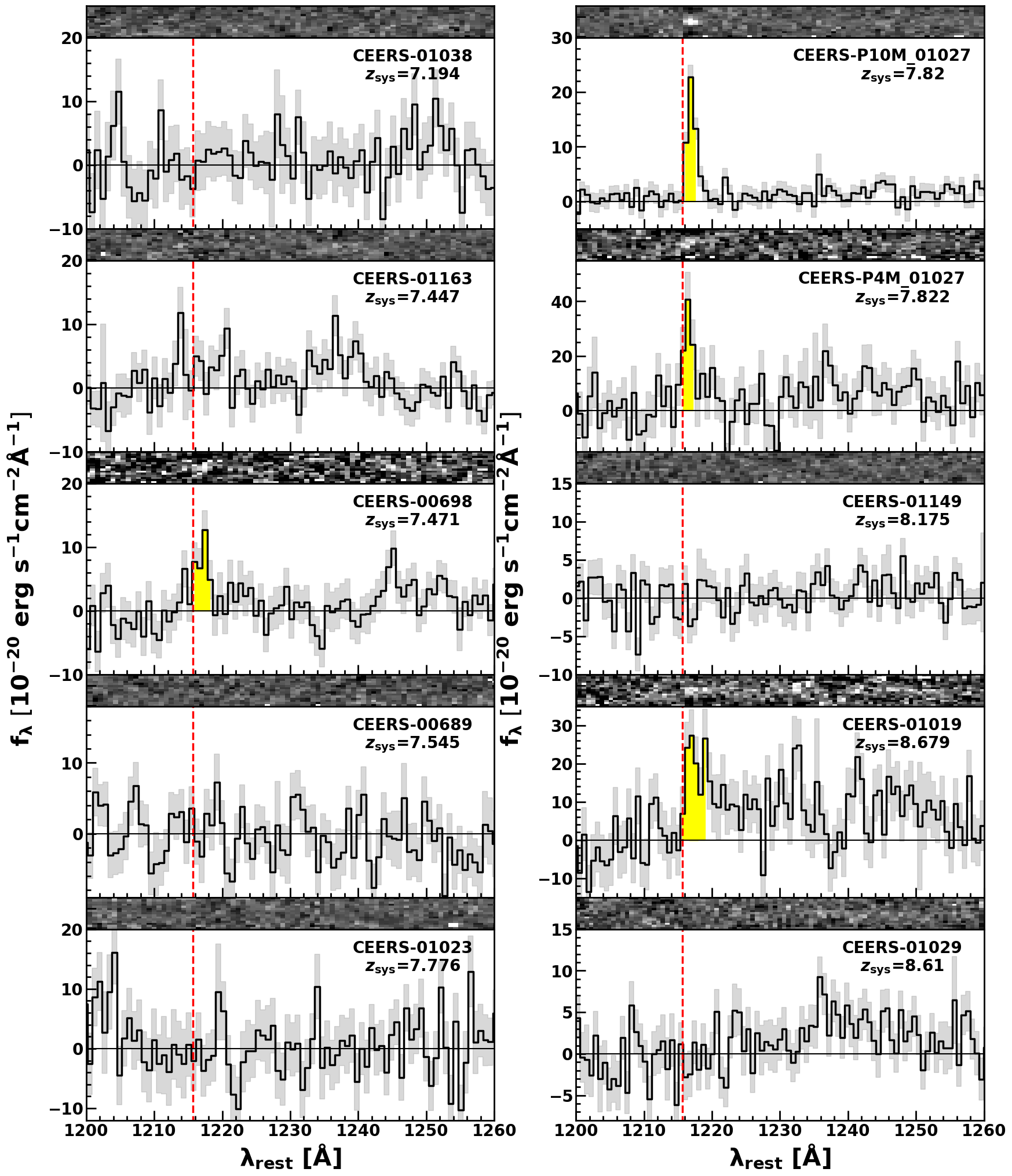}
    \caption{Continued.}
\end{figure*}

\setcounter{figure}{0}
\begin{figure*}[t]
    \centering
    \includegraphics[scale=0.42]{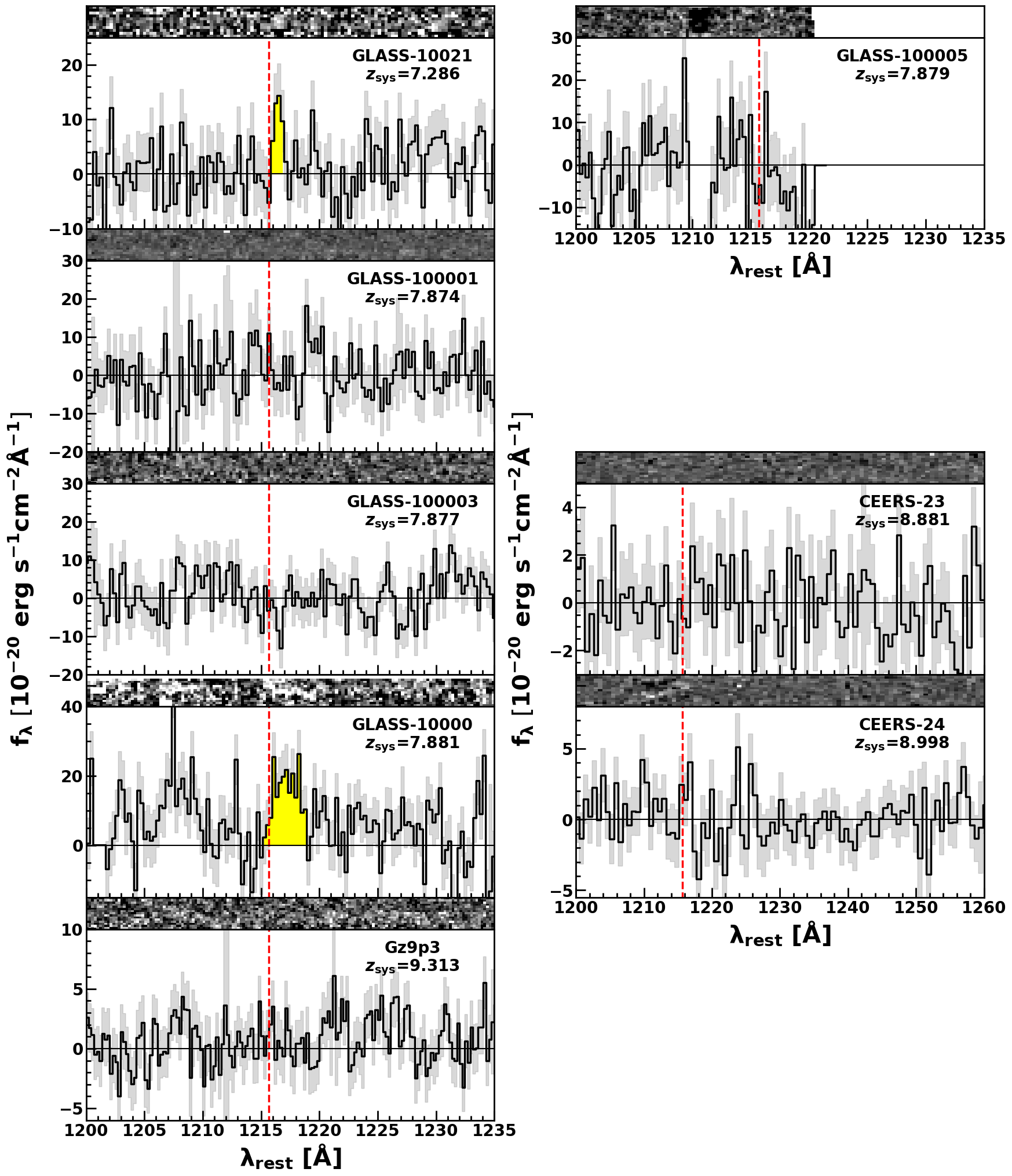}
    \caption{Continued.}
\end{figure*}

\begin{figure*}[t]
    \centering
    \includegraphics[scale=0.42]{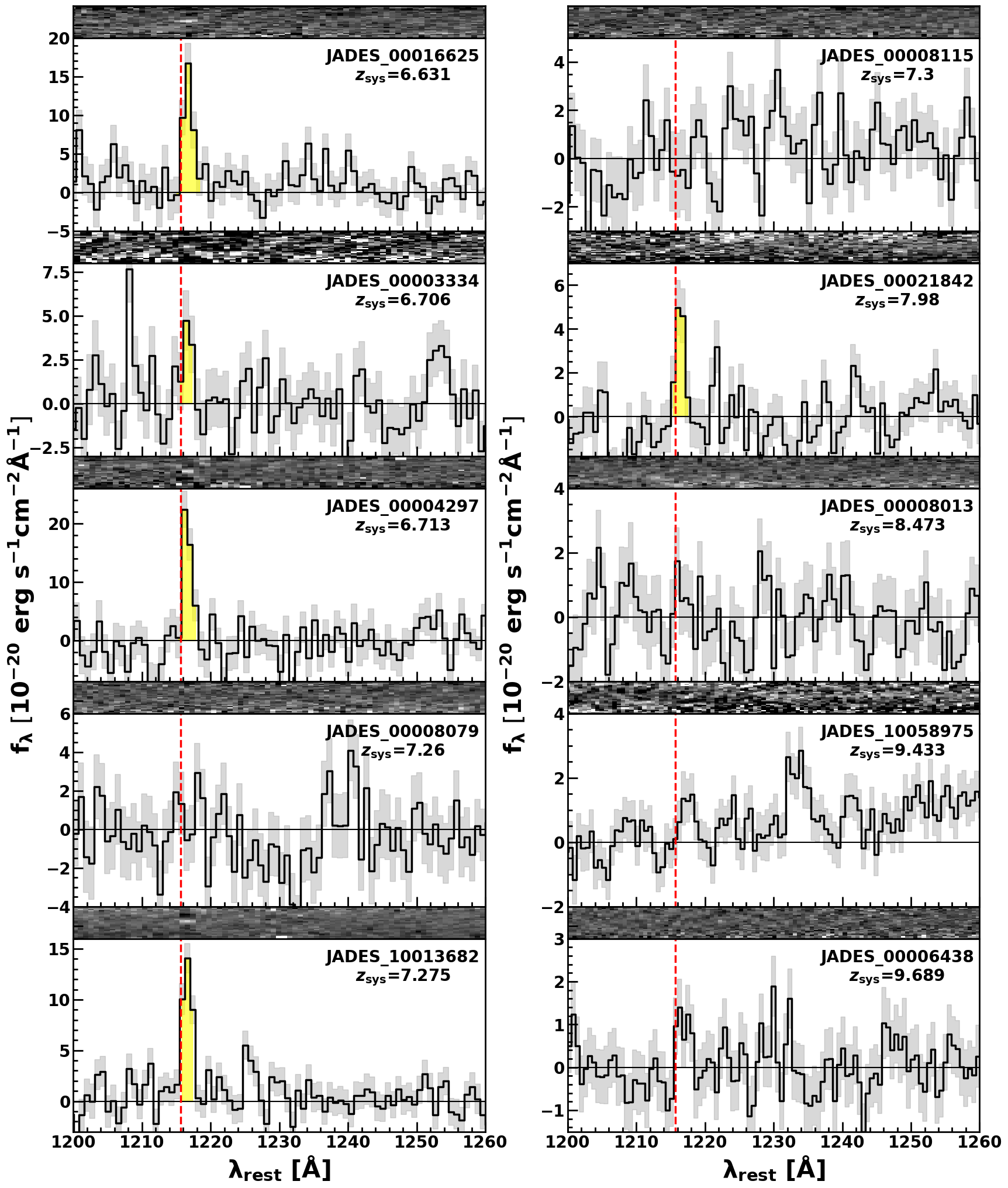}
    \caption{Same as Figure 1, but for the JADES-data sample.}
    \label{fig:JADES-data sample}
\end{figure*}

\setcounter{figure}{1}
\begin{figure*}[t]
    \centering
    \includegraphics[scale=0.42]{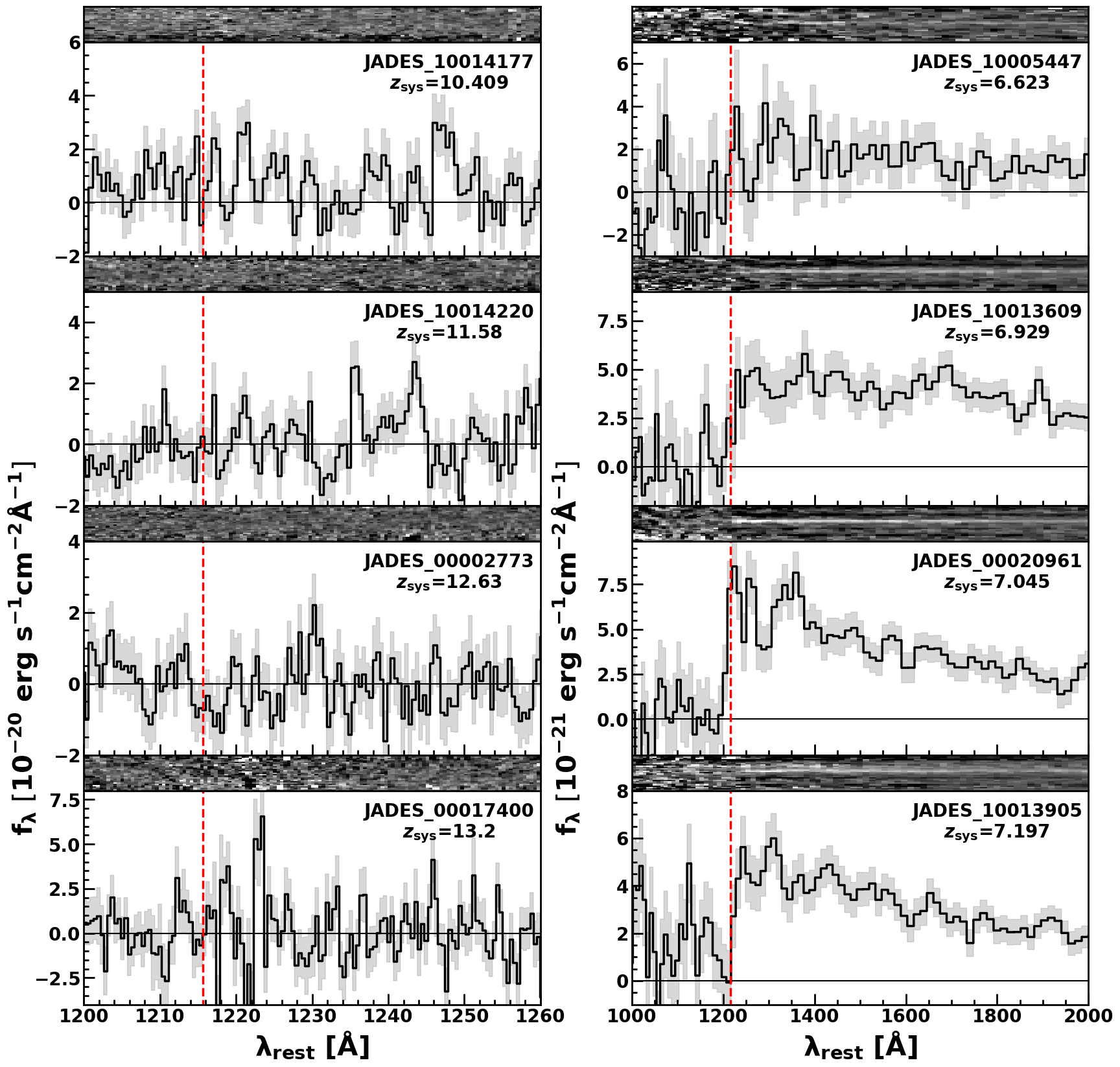}
    \caption{Continued.}
\end{figure*}

\clearpage
\begin{figure}[t]
    \centering
    \includegraphics[scale=0.55]{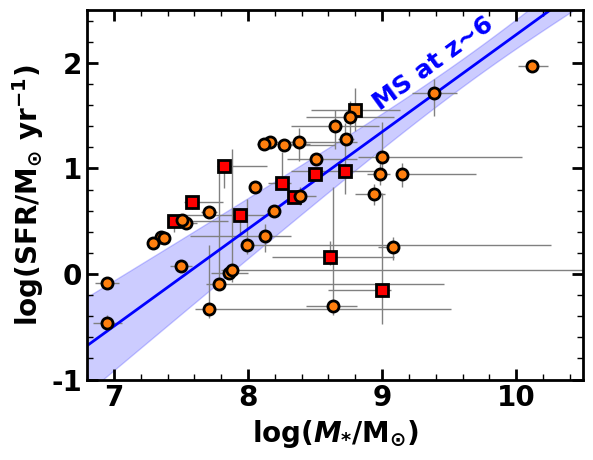}
    \caption{Star formation rate as a  function of stellar mass. The red squares and orange circles present our sample galaxies whose SFRs are measured in this work and the literature \citep{Bunker2023a,Curti2023,Harikane2024,Nakajima2023}, respectively. The blue solid line and shaded region indicate the star-formation main sequence (MS) at $z\sim6$ and its uncertainty \citep{Santini2017}, respectively.}    
    \label{fig:stellar_mass-SFR}
\end{figure}

\noindent We show the spectra of 53 galaxies other than GN-z11 in Figures \ref{fig:public-data_sample} and \ref{fig:JADES-data sample}. We note that because CEERS\_01027 is observed at two pointing positions with the medium-resolution grating, two spectra of CEERS\_01027 (CEERS\_P4M\_01027 and CEERS\_P10M\_01027) are shown in Figure \ref{fig:public-data_sample}.

In Figure \ref{fig:stellar_mass-SFR}, we compare the stellar mass $M_*$ and star formation rate (SFR) of our galaxies with those of the star-formation main sequence to confirm whether our galaxies are typical star-forming galaxies or not. We adopt the stellar mass and SFR reported by \citet{Nakajima2023} and \citet{Harikane2024} for 14 and 5 galaxies of the public-data sample, respectively. We use the stellar mass and SFR reported by \citet{Curti2023}, \citet{Harikane2024}, and \citet{Bunker2023a} for 11, 3, and 1 galaxies of the JADES-data sample, respectively. For 11 galaxies which \citet{Nakajima2023} only report the stellar mass estimates, we calculate the SFR from the UV magnitude. We first derive the UV luminosity $L_\nu(\mathrm{UV})$ from the UV magnitude. We then convert the UV luminosity into the SFR with $\mathrm{SFR}=1.15\times10^{-28}L_\nu(\mathrm{UV})$ \citep{Madau2014}. We correct the SFR for the dust extinction, using an extinction$-$UV slope $\beta_\mathrm{UV}$ relation \citep{Meurer1999} and $\beta_\mathrm{UV}-\mathrm{M}_\mathrm{UV}$ relation \citep{Bouwens2014}. In Figure \ref{fig:stellar_mass-SFR}, we do not present the other 9 galaxies whose stellar mass is not determined. The stellar masses and SFRs of our galaxies are comparable with the star-formation main sequences at $z\sim6$ \citep{Santini2017}, indicating that our galaxies are typical star-forming galaxies.

\section{Measurements of Lyman alpha fluxes and the properties} 
\label{sec:measure}

\subsection{Spectral Fitting}
\label{subsec:fitting}

To obtain the $\Lya$ emission properties of galaxies, we fit a continuum+line model to the $\Lya$ emission line. This model is the linear combination of power-law $a\lambda^{-\beta}$ and Gaussian $A\exp\qty[-\qty(\lambda-\lambda_\mathrm{cen})^2/\sigma^2]$ functions, where $a$, $\beta$, $A$, $\lambda_\mathrm{cen}$, and $\sigma$ are the free parameters. This model is multiplied by $\exp\qty[-\tau(\lambda_\mathrm{obs})]$ where $\tau(\lambda_\mathrm{obs})$ is the optical depth of the IGM to the $\Lya$ photons at the observed wavelength $\lambda_\mathrm{obs}$ \citep{Inoue2014}. We then convolve the model with the line spread functions (LSF) of JWST/NIRSpec measured with a spectrum of a planetary nebula \citep{Isobe2023a} to account for the instrumental smoothing. 

\begin{figure}
    \centering
    \includegraphics[scale=0.3]{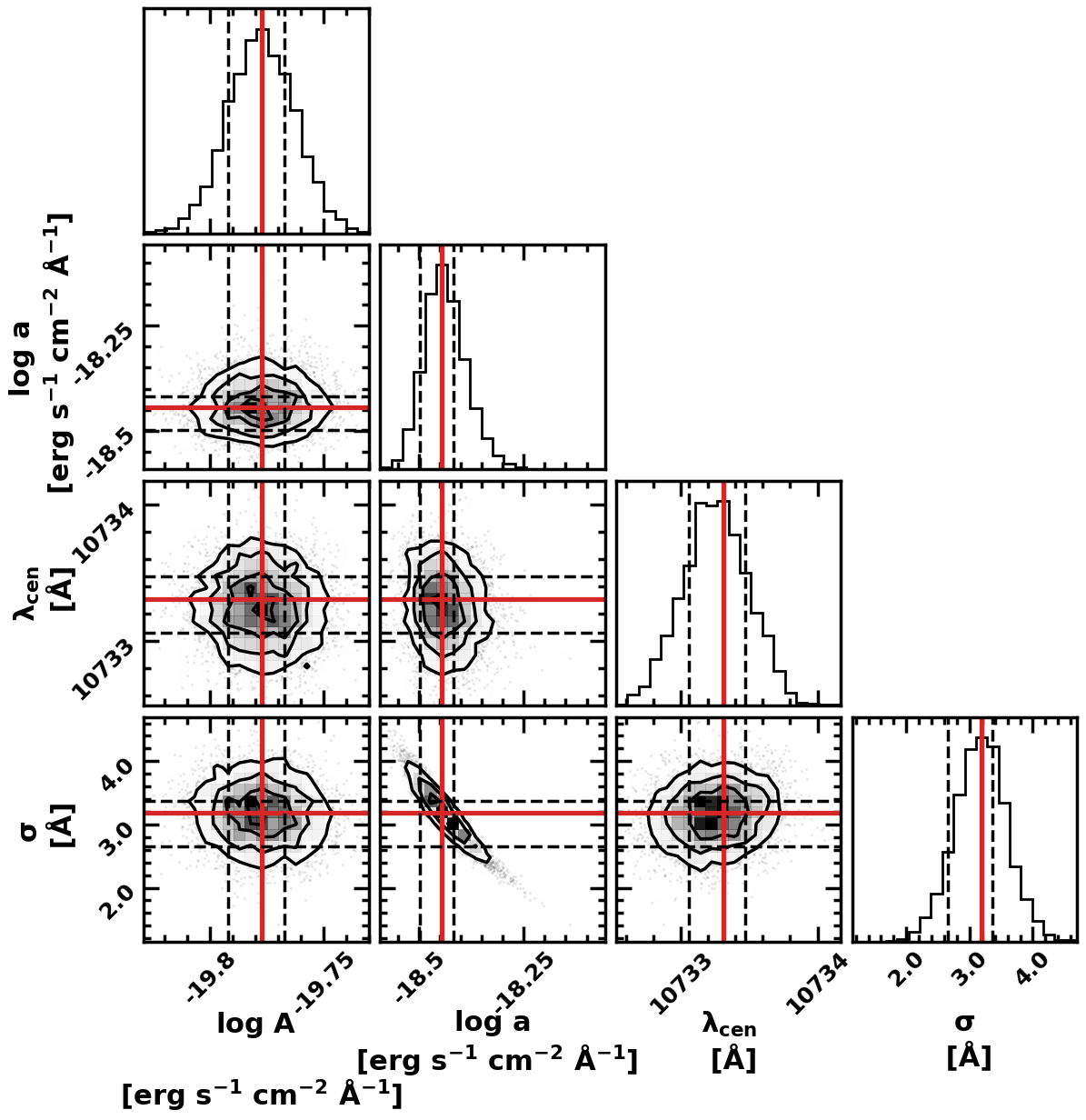}
    \includegraphics[scale=0.3]{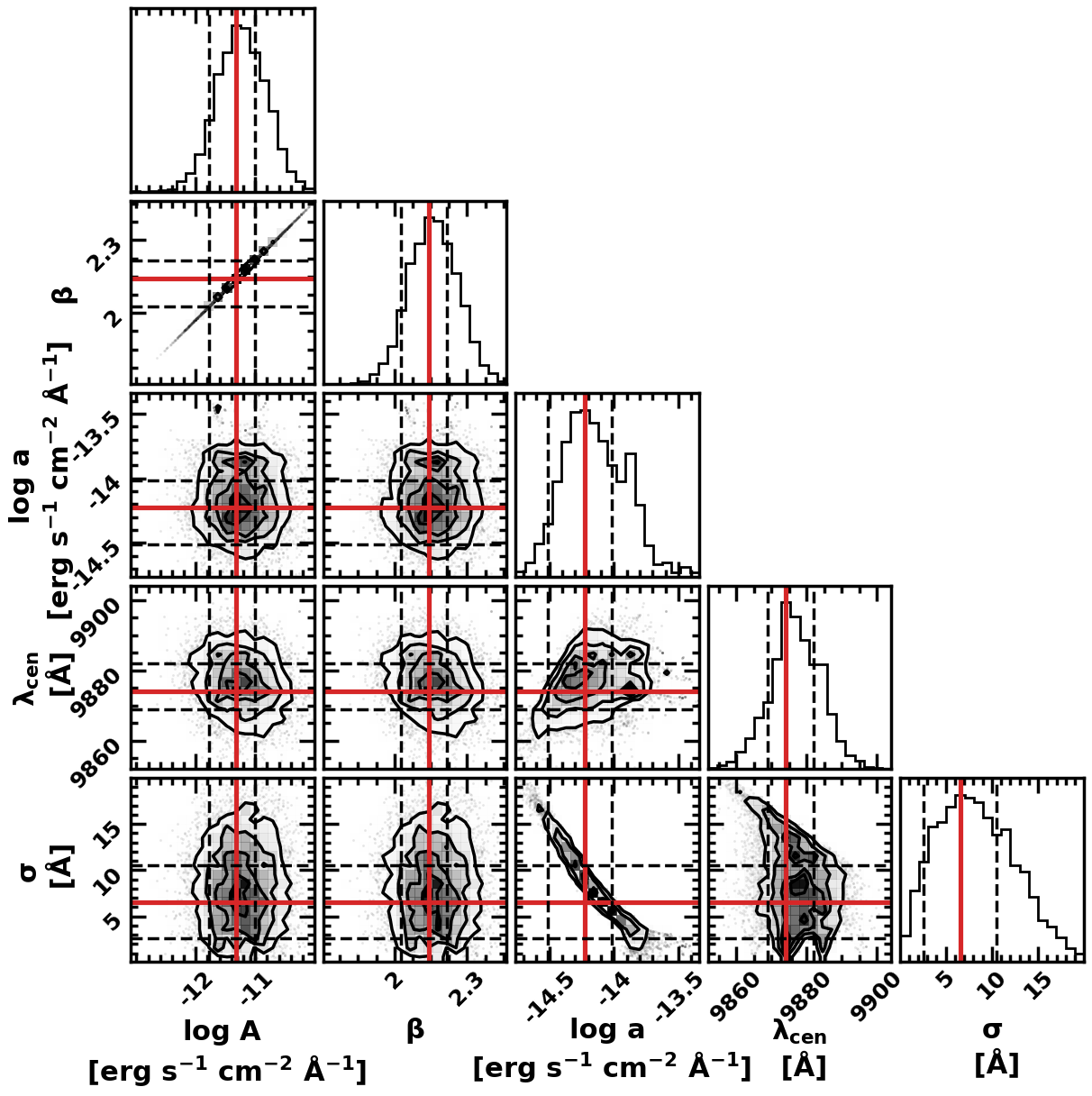}
    \caption{The posterior probability distribution functions of the fitting parameters. The top (bottom) panel represents the fitting results for a grating (prism) spectrum of CEERS-P10M\_01027 (CEERS\_00439) . The red solid lines and black dashed lines show the mode and the boundaries of the 68th percentile highest posterior density intervals (HPDI), respectively (Section \ref{subsec:fitting}).}
    \label{fig:MCMC}
\end{figure}

We determine the free parameters of the model, using \texttt{emcee} \citep{Foreman2013} to conduct Marchov Chain Monte Carlo (MCMC). We apply this fitting method to the prism spectra of galaxies in the public-data sample. For both of the grating and prism spectra of galaxies in the JADES-data sample,  we add the full width half maximum (FWHM) for the LSF as an additional free parameter. Because the spectral resolution of the JADES-data sample is higher than those of the other spectra due to the compact morphologies of the JADES targets \citep{Jones2023}, differences of the FWHM for the LSF have influences on the fittings. The signal to noise ratio of the continua in the grating spectra of our galaxies is worse than that in the prism spectra, which do not allow us to simultaneously determine the amplitude $a$ and the slope $\beta$ of the power law. We thus fix the slope of the power law to zero for the grating spectra under the assumption of a flat continuum ($f_\nu=$const.) that is typical for the star-forming galaxies \citep{Stark2010} and determine the rest of the model parameters. We show examples of the posterior distributions of the fitting parameters in Figure \ref{fig:MCMC}. We determine the model parameter and $1\sigma$ uncertainty by the mode (i.e., a peak of the posterior distribution) and $68$th percentile highest posterior density interval (HPDI; i.e., a narrowest interval containing $68\%$) of the posterior distribution, respectively, which are good representatives even for a skewed posterior distribution. We are not able to conduct spectral fittings to the grating spectra of JADES\_10013905 and GLASS\_100005 due to the small numbers of spectral pixels on the redder side than the observed $\Lya$ wavelengths. As for JADES\_10013905, we use the prism spectrum for the fitting. 

\clearpage

\begin{table*}[t]
    \begin{center}
    \tabcolsep = 1pt
    \caption{Sample in this study}
    \label{tab:all}
    \begin{tabular}{cccccccccc}
         \hline
         \hline
         Name&$z_{\mathrm{sys}}$&$z_{\mathrm{Ly}\alpha}$&$\mathrm{M_{UV}}$&$\EWLya$&$\dvLya$&FWHM&$\log M_*$&$\log \mathrm{SFR}$&Ref. \\
         &&&[mag]&[\AA]&[km $\mathrm{s^{-1}}$]&[km $\mathrm{s^{-1}}$]&[$M_\odot$]&[$M_\odot\ \mathrm{yr^{-1}}$]\\
         (1)&(2)&(3)&(4)&(5)&(6)&(7)&(8)&(9)&(10)\\
         \hline
        JADES\_10005447 & $6.623$ (Line) & -- & $-17.61$ & $<60.9^{\dag}$ & -- & -- &$6.95^{+0.11}_{-0.11}$&$-0.47^{+0.08}_{-0.06}$& Bu23b, Jo23, Cu23 \\
        JADES\_$00016625^{\P}$ & $6.631$ (Line) & $6.637$ & $-18.76$ & $26.6\pm7.1$ & $234\pm31$ & $408\pm110$ &$7.35^{+0.03}_{-0.03}$&$0.35^{+0.02}_{-0.01}$& Bu23b, Jo23, Cu23 \\
        JADES\_$00003334^{\P}$ & $6.706$ (Line) & $6.712$ & $-17.96$ & $16.5\pm4.2$  & $229\pm113$ & $207\pm128$ &$7.50^{+0.08}_{-0.08}$&$0.08^{+0.04}_{-0.04}$& Bu23b, Jo23, Cu23 \\
        JADES\_$00004297^{\P}$ & $6.713$ (Line) & $6.718$ & $-18.51$ & $36.6\pm5.0$ & $188\pm54$ & $447\pm60$ &$7.29^{+0.03}_{-0.03}$&$0.30^{+0.02}_{-0.01}$& Bu23b, Jo23, Cu23\\
        JADES\_10013609 & $6.929$ (Line) & -- & $-18.69$ & $<1.9^{\dag}$ & -- & -- &$7.71^{+0.06}_{-0.06}$&$0.59^{+0.02}_{-0.02}$& Bu23b, Jo23, Cu23 \\
        JADES\_00020961 & $7.045$ (Line) & -- & $-18.84$ & $<12.2^{\dag}$ & -- & -- &$7.86^{+0.14}_{-0.14}$&$0.01^{+0.05}_{-0.04}$& Bu23b, Jo23, Cu23 \\
        CEERS\_00542 & $7.064$ (Line) & -- & $-19.89$ & $<47.6^{\dag}$ & -- & -- &$8.25^{+0.07}_{-0.10}$&$0.86^{+0.30}_{-0.05}$& Na23, This \\
        CEERS\_00044 & $7.104$ (Line) & -- & $-19.38$  & $62.6\pm19.5^{\dag}$ & -- & -- &$7.94^{+0.28}_{-0.42}$&$0.56^{+0.07}_{-0.17}$&Na23, This \\
        CEERS\_00534 & $7.115$ (Line) & -- & $-20.10$ & $<35.3^{\dag}$  & -- & -- &$8.72^{+0.28}_{-0.40}$&$0.98^{+0.46}_{-0.22}$&Na23, This \\
        CEERS\_00829 & $7.167$ (Line) & -- & $-19.57$ & $<41.3^{\dag}$  & -- & -- & $7.58^{+0.23}_{-0.06}$&$0.68^{+0.04}_{-0.06}$&Na23, This \\
        CEERS\_80374 & $7.174$ (Line) & -- & $-18.09$  & $<86.7^{\dag}$ & -- & -- &$9.00^{+0.07}_{-0.40}$&$-0.15^{+0.91}_{-0.33}$& Na23, This \\
        CEERS\_00439 & $7.179$ (Line) & -- & $-19.28$  & $33.8\pm8.0^{\dag}$ & -- & -- &$7.45^{+0.40}_{-0.05}$&$0.50^{+0.04}_{-0.10}$& Na23, This \\
        CEERS\_00498 & $7.179$ (Line) & -- & $-20.20$ & $<30.6^{\dag}$ & -- & -- &$7.82^{+0.32}_{-0.05}$&$1.02^{+0.05}_{-0.21}$& Na23, This \\
        CEERS\_01038 & $7.194$ (Line) & -- & $-19.24$ & $<9.1$ & -- & -- &$8.39^{+0.12}_{-0.11}$&$0.74^{+0.08}_{-0.09}$& Na23  \\
        JADES\_10013905 & $7.197$ (Line) & -- & $-18.80$ & $<7.2^{\dag*}$  & -- & -- &$7.37^{+0.05}_{-0.05}$&$0.34^{+0.03}_{-0.02}$& Bu23b, Cu23 \\
        JADES\_00008079 & $7.260$ (Line) & -- & $-17.95$ & $<8.2$ & -- & -- &$-$&$-$& Bu23b, Jo23  \\
        JADES\_$10013682^{\P}$ & $7.275$ (Line) & $7.281$ & $-17.00$ & $31.5\pm3.3$ & $215\pm23$ & $261\pm57$ &$6.95^{+0.09}_{-0.09}$&$-0.09^{+0.04}_{-0.04}$&Bu23b, Jo23, Cu23\\
        GLASS\_10021 & $7.286$ (Line) & $7.292$ & $-21.44$ & $3.2\pm1.0$ &$203\pm32$ & $121\pm72$ &$8.51^{+0.30}_{-0.22}$&$1.09^{+0.02}_{-0.02}$& Na23 \\
        JADES\_00008115 & $7.3$ (Break) & -- & -- & $<4.3$ & -- & -- &$-$&$-$& Bu23b  \\
        CEERS\_01163 & $7.455$ (Line) & -- & $>-20.78$ & $<9.3$ & -- & -- &$<8.98$&$-$& Na23  \\
        CEERS\_$00698^{\P}$ & $7.471$ (Line) & $7.480$ & $-21.60$  & $5.4\pm1.0$ &$334\pm64$& $354\pm135$ &$9.39^{+0.17}_{-0.17}$&$1.71^{+0.14}_{-0.21}$&Na23 \\
        CEERS\_80432 & $7.477$ (Line) & -- & $-20.05$ & $<27.3^{\dag}$ & -- & -- &$8.50^{+0.05}_{-0.05}$&$0.95^{+0.03}_{-0.02}$& Na23, This  \\
        CEERS\_80372 & $7.483$ (Line) & -- & $-19.27$ & $<34.5^{\dag}$ & -- & -- &$8.19^{+0.05}_{-0.05}$&$0.60^{+0.05}_{-0.06}$& Na23  \\
        CEERS\_80239 & $7.487$ (Line) & -- & $-17.92$ & $105.3\pm24.0^{\dag}$ & -- & -- &$8.13^{+0.19}_{-0.56}$&$0.36^{+0.11}_{-0.15}$& Na23 \\
        CEERS\_80445 & $7.509$ (Line) & -- & $-19.66$ & $<18.2^{\dag}$ & -- & -- &$8.34^{+0.05}_{-0.05}$&$0.73^{+0.02}_{-0.01}$& Na23, This  \\
        CEERS\_00689 & $7.552$ (Line) & -- & $-21.98$ & $<8.3$ & -- & -- &$<8.70$&$-$& Na23, Ju23  \\
        CEERS\_00686 & $7.752$ (Line) & -- & $-20.68$ & $20.4\pm6.6^{\dag}$& -- & -- &$<8.44$&$-$&Na23, Ju23   \\
        CEERS\_01023 & $7.776$ (Line) & -- & $-21.06$ & $<17.5$ & -- & -- &$8.76^{+0.33}_{-0.33}$&$1.49^{+0.15}_{-0.22}$& Na23  \\
        CEERS\_$01027^{\ddag\P}$ & $7.821$ (Line) & $7.828$ & $-20.73$  & $17.9\pm2.5$ & $232\pm56$ & $283\pm57$ &$8.12^{+0.34}_{-0.05}$&$1.23^{+0.03}_{-0.03}$& Na23  \\
        GLASS\_100001 & $7.874$ (Line) & -- & $-20.29$ & $<16.7$ & -- & -- &$9.15^{+0.55}_{-0.00}$&$0.95^{+0.10}_{-0.13}$& Na23  \\
        GLASS\_100003 & $7.877$ (Line) & -- & $-20.69$ & $<0.2$ & -- & -- &$8.27^{+0.38}_{-0.04}$&$1.22^{+0.04}_{-0.05}$& Na23  \\
        GLASS\_100005 & $7.879$ (Line) & -- & $-20.06$ & $-^{*}$ & -- & -- &$8.38^{+0.43}_{-0.13}$&$1.25^{+0.13}_{-0.18}$& Na23  \\
        GLASS\_10000 & $7.881$ (Line) & $7.890$ & $-20.36$ & $7.5\pm1.3$ & $308\pm102$ & $818\pm250$ &$8.16^{+0.27}_{-0.07}$&$1.25^{+0.03}_{-0.04}$& Na23\\
        JADES\_$00021842^{\P}$ & $7.980$ (Line) & $7.985$ & $-18.71$ & $18.8\pm4.9$  & $168\pm91$ & $277\pm115$ &$7.51^{+0.04}_{-0.04}$&$0.51^{+0.03}_{-0.03}$&Bu23b, Jo23, Cu23\\
        CEERS\_00003 & $8.005$ (Line) & -- & $-18.76$& $<186.3^{\dag}$ & -- & -- &$8.61^{+0.48}_{-0.43}$&$0.16^{+0.15}_{-0.08}$& Na23, This  \\
        CEERS\_01149 & $8.184$ (Line) & -- & $-20.83$ & $<11.8$ & -- & -- &$8.65^{+0.33}_{-0.33}$&$1.40^{+0.15}_{-0.22}$& Na23  \\
        JADES\_00008013 & $8.473$ (Line) & -- & $-17.54$ &$<6.0$ & -- & -- &$7.54^{+0.08}_{-0.08}$&$0.48^{+0.05}_{-0.04}$& Bu23b, Jo23, Cu23  \\
        CEERS\_01029 & $8.615$ (Line) & -- & $-21.14$  & $<9.8$ & -- & -- &$8.80^{+0.33}_{-0.33}$&$1.56^{+0.20}_{-0.20}$& Na23, This  \\
        CEERS\_90671 & $8.637$ (Line) & -- & $-18.7$ & $<45.2^{\dag}$ & -- & -- &$8.94^{+0.08}_{-0.14}$&$0.76^{+0.09}_{-0.11}$& AH23b, Ha24, Na23  \\
        CEERS\_01019 & $8.679$ (Line) & $8.686$ & $-22.44$  & $3.4\pm1.1$ &$231\pm54$ & $978\pm248$ &$10.12^{+0.12}_{-0.11}$&$1.97^{+0.04}_{-0.05}$&Zi15, Ha24, La23, \\
        &&&&&&&&&Na23, Sa23, Ta23 \\
        CEERS1\_6059 & $8.873$ (Line) & -- & $-20.75$ & $<14.3^{\dag}$ & -- & -- &$8.98^{+0.08}_{-0.10}$&$0.94^{+0.12}_{-0.10}$& Fu23, Ha24, Na23 \\
        CEERS-23 & $8.881$ (Line) & -- & $-18.9$ & $<21.7$ & -- & -- &$7.88^{+2.88}_{-0.09}$&$0.04^{+1.14}_{-0.12}$& Fu23, Ha24, Ta23  \\
        CEERS-24 & $8.998$ (Line) & -- & $-19.4$ & $<22.7$ & -- & -- &$7.99^{+2.27}_{-0.00}$&$0.28^{+0.43}_{-0.09}$& Fu23, Ha24, Ta23  \\
        Gz9p3 & $9.313$ (Line) & -- & $-21.6$ & $<2.3$ & -- & -- &$-$&$-$& Bo23, Ha24  \\
        JADES\_10058975 & $9.433$ (Line) & -- & $-20.36$ & $<0.7$  & -- & -- &$8.05^{+0.03}_{-0.03}$&$0.82^{+0.01}_{-0.01}$& Bu23b, Jo23, Cu23 \\
        JADES\_00006438 & $9.689$ (Line) & -- & $-19.27$ & $<17.7$  & -- & -- &$-$&$-$& Bu23b, Jo23 \\
        MACS0647-JD & $10.17$ (Line) & -- & $-20.3$ & $<5.9^{\dag}$ & -- & -- &$-$&$-$& Ha24, Hs23 \\
        JADES\_10014177 & $10.409$ (Break) & -- & $-18.45$ & $<4.2$ & -- & -- &$-$&$-$& Jo23  \\
        GN-z11 & $10.603$ (Line) & $10.624$  & $-21.5$ & $18.0\pm2.0$ & $555\pm32$ & $566\pm61$ &$8.73^{+0.06}_{-0.06}$&$1.27^{+0.02}_{-0.02}$&Bu23a  \\
        \hline
    \end{tabular}
    \end{center}
\end{table*}

\setcounter{table}{0} 
\begin{table*}[t]
    \begin{center}
    \caption{Sample of this study (Continued)}
    \begin{tabular}{cccccccccc}
    \hline
    \hline
    Name&$z_{\mathrm{sys}}$&$z_{\mathrm{Ly}\alpha}$&$\mathrm{M_{UV}}$&$\EWLya$&$\dvLya$&FWHM&$\log M_*$&$\log \mathrm{SFR}$& Ref. \\
    &&&[mag]&[\AA]&[km $\mathrm{s^{-1}}$]&[km $\mathrm{s^{-1}}$]&[$M_\odot$]&[$M_\odot\ \mathrm{yr^{-1}}$]&\\
    (1)&(2)&(3)&(4)&(5)&(6)&(7)&(8)&(9)&(10)\\
    \hline
    CEERS2\_588 & $11.04$ (Line) & -- & $-20.4$ & $<11.5^{\dag}$ & -- & -- &$9.00^{+1.05}_{-0.18}$&$1.10^{+0.33}_{-0.17}$& Ha24  \\
    CR2-z12-1 & $11.40$ (Line) & -- & $-20.1$ & $<9.4^{\dag}$ & -- & -- &$7.79^{+1.67}_{-0.10}$&$-0.10^{+1.14}_{-0.05}$& AH23a, Ha24  \\
    JADES\_10014220 & $11.58$ (Break) & -- & $-19.36$ & $<2.3$ & -- & -- &$9.08^{+0.04}_{-0.11}$&$0.26^{+0.10}_{-0.12}$& Bu23b, CL23, Ha24  \\
    JADES\_00002773 & $12.63$ (Break) & -- & $-18.8$ & $<5.8$ & -- & -- &$8.63^{+0.18}_{-0.20}$&$-0.30^{+1.13}_{-0.09}$& Bu23b, CL23, Ha24 \\
    JADES\_00017400 & $13.20$ (Break) & -- & $-18.5$ & $<3.8$  & -- & -- &$7.71^{+1.81}_{-0.10}$&$-0.30^{+0.61}_{-0.09}$& Bu23b, CL23, Ha24 \\
    \hline
    \end{tabular}
    \end{center}
    \tablecomments{(1): Name. (2): Systemic redshift. The spectroscopic feature used to determine the redshift is noted (Break: Lyman break, Line: emission line). (3): $\Lya$ redshift. (4): Absolute UV magnitude. (5): Rest-frame $\Lya$ equivalent width and the $1\sigma$ error. For galaxies with no $\Lya$ detections, we show $3\sigma$ upper limits. (6): $\Lya$ velocity offset. (7): Full width half maximum of the $\Lya$ emission line. (8): Stellar mass. (9): Star-formation rate. (10): References for systemic redshifts, absolute UV magnitude, stellar masses, and SFRs. (This: this work, AH23a: \citealt{Haro2023a}, AH23b: \citealt{Haro2023b}, Bo23: \citealt{Boyett2023}, Bu23a: \citealt{Bunker2023a}, Bu23b: \citealt{Bunker2023b}, CL23: \citealt{Curtis-Lake2023}, Cu23: \citealt{Curti2023}, Fu23: \citealt{Fujimoto2023}, Ha24: \citealt{Harikane2024}, Hs23: \citealt{Hsiao2023}, Jo23: \citealt{Jones2023},  Ju23: \citealt{Jung2023},  La23: \citealt{Larson2023}, Na23: \citealt{Nakajima2023}, Sa23: \citealt{Sanders2023}, Ta23: \citealt{Tang2023}, Zi15: \citealt{Zitrin2015}). \\ $^\dag$ The $\EWLya$ values and $3\sigma$ upper limits of these galaxies are measured with the prism spectra. \\ $^*$ Because the numbers of spectral pixels on the redder side than the observed $\Lya$ wavelengths to determine the continuum fluxes are small for the grating spectra, we are not able to determine the upper limits of $\Lya$ EW. As for JADES\_10013905, we adopt the measurement from the prism spectrum. \\ $^\ddag$ The systemic redshift, $\Lya$ redshift, EW, velocity offset, and FWHM presented here are mean values of those measured from two spectra obtained at different pointing positions. \\ $^\P$ These galaxies are used to construct the composite spectra (Section \ref{subsec:profile}).}
\end{table*}

\begin{longrotatetable}
\begin{table*}[t]
    \begin{center}
    \tabcolsep = 1pt
    \caption{Comparison with the previous studies.}
    \label{tab:comparison}
     \begin{tabular}{ccccccccccccc}
        \hline
        \hline
        ID&$z_\mathrm{sys}$&$z_\mathrm{Ly\alpha}$&$\EWLya$&$\mathrm{EW}_\mathrm{Ly\alpha, Ta23}$&$\mathrm{EW}_\mathrm{Ly\alpha, Jo23}$&$\mathrm{EW}_\mathrm{Ly\alpha, Sa23}$&$\mathrm{EW}_\mathrm{Ly\alpha, Ch23}$&$\dvLya$&$\Delta v_\mathrm{Ly\alpha, Ta23}$&$\Delta v_\mathrm{Ly\alpha, Jo23}$&$\Delta v_\mathrm{Ly\alpha, Sa23}$&$\Delta v_\mathrm{Ly\alpha, Ch23}$\\
        &&&[\AA]&[\AA]&[\AA]&[\AA]&[\AA]&[km $\mathrm{s^{-1}}$]&[km $\mathrm{s^{-1}}$]&[km $\mathrm{s^{-1}}$]&[km $\mathrm{s^{-1}}$]&[km $\mathrm{s^{-1}}$]\\
        (1)&(2)&(3)&(4)&(5)&(6)&(7)&(8)&(9)&(10)&(11)&(12)&(13)\\
        \hline
         JADES\_$00016625^\ddag$ & 6.631 & 6.637 & $26.6\pm7.1$ & -- & $38.9\pm17.1$ & $51.0\pm7.4$ & -- & $234\pm31$ & --  & -- & $244.2\pm50.9$& -- \\
         JADES\_00003334 & 6.706 & 6.712 & $16.5\pm4.2$ & -- & -- & -- & -- & $229\pm113$ & --  & -- & --& -- \\
         JADES\_00004297 & 6.713 & 6.718 & $36.6\pm5.0$ & -- & -- & -- & -- & $188\pm54$ & -- & -- & --& -- \\
         CEERS\_00044 & 7.104 & -- & $62.6\pm19.5^{\dag}$ & $77.6\pm5.5^{\dag}$ & -- & -- & -- & -- & -- & -- & --& -- \\
         CEERS\_$80374^\ddag$ & 7.174 & -- & $<86.71^{\dag}$ & -- & -- & -- & $205^{+48\dag}_{-27}$ & -- & -- & -- & --& -- \\
         CEERS\_00439 & 7.179 & -- & $33.8\pm8.0^{\dag}$ & -- & -- & -- & -- & -- & -- & -- & --& -- \\
         JADES\_$10013682^\P$ & 7.275 & 7.281 & $31.5\pm3.3$ & -- & $147.8\pm19.6$ & $337.2\pm175.5$ & -- & $215\pm23$ & -- & -- & $178.4\pm48.9$& -- \\
         GLASS\_10021 & 7.286 & 7.292 & $3.2\pm1.0$ & -- & -- & -- & -- & $203\pm32$ & -- & -- & --& -- \\
         CEERS\_00698 & 7.471 & 7.480 & $5.4\pm1.0$ & $9.5\pm3.1$ & -- & -- & -- & $334\pm64$ & $545\pm184$ & -- & --& -- \\
         CEERS\_$80239^\ddag$ & 7.487 & -- & $105.3\pm24.0^{\dag}$ & -- & -- & -- & $334^{+109\dag}_{-62}$ & -- & -- & -- & --& -- \\
         CEERS\_00686 & 7.752 & -- & $20.4\pm6.6^{\dag}$ & $41.9\pm1.6^{\dag}$ & -- & -- & -- & -- & -- & -- & -- & -- \\
         CEERS\_01027 & 7.821 & 7.828 & $17.9\pm2.5$ & $20.4\pm3.1$ & -- & -- & -- & $232\pm56$ & $323\pm18$ & -- & -- & -- \\
         GLASS\_10000 & 7.881 & 7.890 & $7.5\pm1.3$ & -- & -- & -- & -- & $308\pm102$ & -- & -- & -- & -- \\
         JADES\_00021842 & 7.98 & 7.985 & $18.8\pm4.9$ & -- & $23.9\pm7.2$ & $29.2\pm3.3$ & -- & $168\pm91$ & -- & -- & $166.5\pm81.3$ & -- \\
         CEERS\_$01029^*$ & 8.615 & -- & $<9.8$ & $4.2\pm1.3$ & -- & -- & -- & -- & $1938\pm162$ & -- & -- & -- \\
         CEERS\_01019 & 8.679 & 8.686 & $3.4\pm1.1$ & $9.8\pm2.4$ & -- & -- & -- & $231\pm54$ & $458\pm161$ & -- & -- & -- \\
        \hline
     \end{tabular}
     \end{center}
     \tablecomments{(1): Name. (2): Systemic redshift. (3): $\Lya$ redshift. (4): Rest-frame $\Lya$ equivalent width measured in this study. (5): Rest-frame $\Lya$ equivalent width measured in \citet{Tang2023}. (6): Rest-frame $\Lya$ equivalent width measured in \citet{Jones2023}. (7): Rest-frame $\Lya$ equivalent width measured in \citet{Saxena2023}. (8): Rest-frame $\Lya$ equivalent width measured in \citet{Chen2024}. (9): $\Lya$ velocity offset measured in this study. (10): $\Lya$ velocity offset measured in \citet{Tang2023}. (11): $\Lya$ velocity offset measured in \citet{Jones2023}. (12): $\Lya$ velocity offset measured in \citet{Saxena2023}. (13): $\Lya$ velocity offset measured in \citet{Chen2024}.\\ $^\dag$ The $\EWLya$ values and $3\sigma$ upper limits of these galaxies are measured with the prism spectra. \\ $^*$ While \citet{Tang2023} claim $\Lya$ detection for this galaxy, we do not detect $\Lya$ emission beyond our threshold of the signal to noise ratio of $>3\sigma$. The $\EWLya$ value of \citet{Tang2023} are within $3\sigma$ upper limits of $\EWLya$ in this work. \\ $^\ddag$ Although \citet{Chen2024} show high $\EWLya$ values for these galaxies, we do not detect $\Lya$ emission for CEERS\_80374 and obtain lower $\EWLya$ for CEERS\_80239. The deviations of the measurements may come from the aperture size to extract the spectra of these galaxies. \\ $^\P$ The lower $\EWLya$ in this work than that of \citet{Jones2023} and \citet{Saxena2023} is due to the continuum flux measured from the spectra with different resolutions, the prism and grating spectra.}
\end{table*}
\end{longrotatetable}

\clearpage
\noindent In summary, we conduct spectral fittings to the grating spectra of $30$ $(=33-2-1)$ galaxies (not including GNz-11) and the prism spectra of $22$ $(=21+1)$ galaxies (see Section \ref{sec:datasam}). Note that we have detected continua for $52$ $(=30+22)$ galaxies including tentative detections. We confirm that if we use galaxies with continua detected at $>3\sigma$ level for the analysis below, our results does not change largely.

With the public-data and JADES-data samples, we find that 11 out of 31 galaxies with the grating spectra and 4 out of 22 galaxies with the prism spectra have $\Lya$ detections, where we place a threshold of the signal to noise ratio of $>3\sigma$ for $\Lya$ detection. In total, we find $15$ $(=11+4)$ out of 53 galaxies with $\Lya$ detections in our sample. We note that $\Lya$ emission of CEERS\_01029 are not detected beyond our threshold, while \citet{Tang2023} claim a detection of $\Lya$ emission. However, we confirm that the $\Lya$ EW measurements of \citet{Tang2023}, $\EWLya=4.2\pm1.3$ \AA, is within the $3\sigma$ upper limits of the $\Lya$ EW in this work, $9.8$ \AA.

Table \ref{tab:all} summarizes the properties of all 54 galaxies in our sample. Table \ref{tab:comparison} compares the rest-frame $\Lya$ equivalent width and $\Lya$ velocity offset of the galaxies with $\Lya$ detections measured in this study with those in the literature. For CEERS\_01027, we show mean values of the $\Lya$ redshift, EW, velocity offset, and line width measured from two grating spectra (Section \ref{sec:datasam}) because the measurements from the two spectra are consistent within errors.

\subsection{Ly$\alpha$ Velocity Offset} 
\label{subsec:vLya}
\begin{figure}[t]
    \centering
    \includegraphics[scale=0.45]{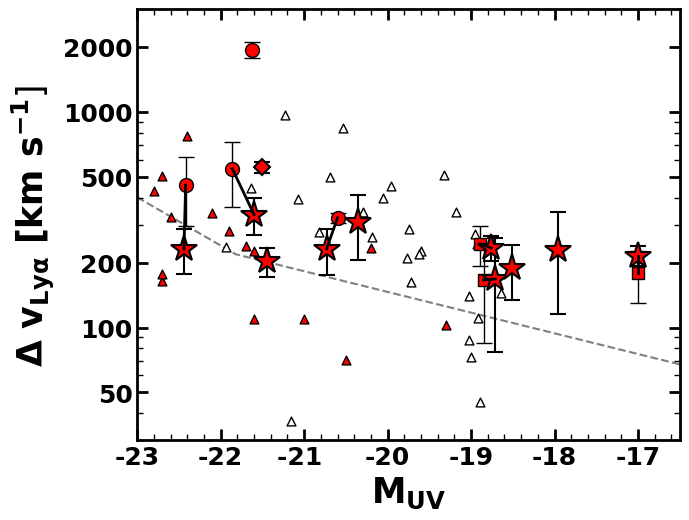}
    \caption{Ly$\alpha$ velocity offset as a function of absolute UV magnitude. The red star marks and diamond show our galaxies and GN-z11 at $z=10.6$ \citep{Bunker2023a} which is included in our sample, respectively.
    The white and red triangles indicate galaxies identified with ground-based telescopes at $z\sim2-3$ \citep{Erb2014} and $z>6$ (\citealt{Cuby2003,Pentericci2011,Pentericci2016,Pentericci2018,Vanzella2011,Willott2013,Willott2015,Maiolino2015,Oesch2015,Stark2015,Stark2017,Furusawa2016,Knudsen2016,Carniani2017,Laporte2017,Mainali2017,Hashimoto2019,Endsley2022}; see \citealt{Endsley2022} and Table 4 therein), respectively. The red circles and squares denote the measurements of \citet{Tang2023} for the CEERS galaxies and \citet{Saxena2023} for the JADES galaxies, respectively, all of which are included in our sample. For the same galaxies, our measurements (star marks) are connected to the \citeauthor{Tang2023}'s measurements (triangles) or \citeauthor{Saxena2023}'s measurements (squares) with the black solid lines. The grey dashed lines represent the empirical relation at $z=7$ suggested by \citet{Mason2018}.}
    \label{fig:dvLya}
\end{figure}

For the galaxies with $\Lya$ emissions detected from the grating spectra, we calculate $\Lya$ redshifts $z_{\Lya}$ with $1+z_{\mathrm{Ly\alpha}}=\lambda_\mathrm{cen}/\lambda_\mathrm{\Lya}$, where $\lambda_\mathrm{\Lya}$ is the rest-frame $\Lya$ wavelength, $1215.67$ \AA.
We then derive the Ly$\alpha$ velocity offset $\dvLya$ that is defined by
\begin{align}
\dvLya=\frac{c(z_{\mathrm{Ly\alpha}}-z_{\mathrm{sys}})}{1+z_{\mathrm{sys}}}
\end{align}
where $c$ is the speed of light. In Figure \ref{fig:dvLya}, we plot the $\dvLya$ values measured in this study and the literature. The $\dvLya$ values of our galaxy are comparable with not only $z>6$ galaxies, but also $z\sim2-3$ galaxies. We confirm that there is no significant redshift evolution of the $\Lya$ velocity offset. For the same galaxies, we compare our measurements with those of \citet{Saxena2023} and \citet{Tang2023} (see also Table \ref{tab:comparison}). The measurements of $\dvLya$ for JADES\_00016625, JADES\_10013682, and JADES\_00021842 are consistent with those in the \citet{Saxena2023}. While the measurements of $\dvLya$ for CEERS\_00698, CEERS\_01027, and CEERS\_01019 appears to be lower than those of \citet{Tang2023} in Figure \ref{fig:dvLya}, they are comparable, considering the slightly lower systemic redshifts reported by \citet{Tang2023} and uncertainties of measurements (Table \ref{tab:comparison}). The range and mean value of the velocity offset in our sample are $168-334$ km $\mathrm{s}^{-1}$ and  $234\pm76$ km $\mathrm{s}^{-1}$, respectively.

\begin{figure}[t]
    \centering
    \includegraphics[scale=0.5]{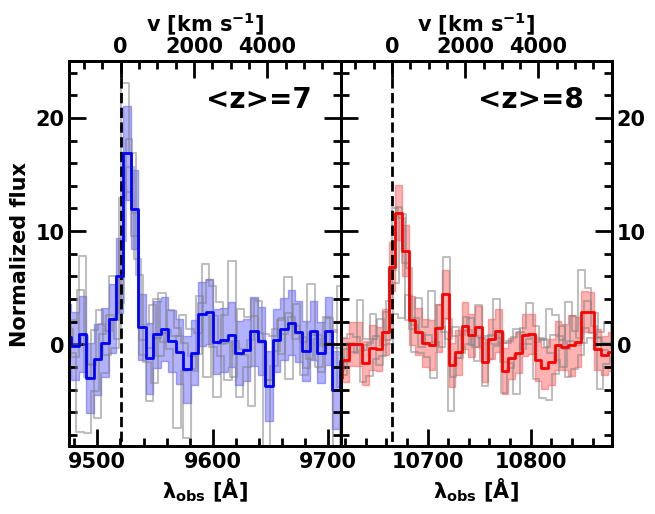}
    \caption{$\Lya$ emission lines in the composite spectra of our sample galaxies. Left: the top and bottom axes are the velocity offset and observed wavelength. The y-axis corresponds to the fluxes normalized with the continuum fluxes. The blue solid line and shaded regions present the composite spectrum for the $\meanz=7$ galaxies and associated $1\sigma$ errors, respectively. The grey solid lines show the individual spectra whose $\Lya$ lines are redshifted to the mean redshift. The vertical black dashed line denotes the $\Lya$ wavelength in the observed frame defined with the mean redshift. Right: same as the left panel, but for the $\meanz=8$ galaxies. The red solid line indicates the composite spectrum for the $\meanz=8$ galaxies.}
    \label{fig:composite}
\end{figure}

\subsection{Ly$\alpha$ Line Profile}
\label{subsec:profile}

In order to investigate $\Lya$ line profile of our galaxies, we make composite spectra by conducting mean stacking for $4$ $(3)$ galaxies at $z\sim7$ $(8)$ that have the medium-resolution grating spectra. We determine the redshift bin so that the numbers of the galaxies at each bin are comparable. The redshift range and mean redshift of the galaxies at $z\sim7$ $(8)$ are $6.6<z<7.3$ $(7.4<z<8.0)$ and $\meanz=7$ $(8)$, respectively. Note that at $\meanz=8$, we stack $4$ spectra for $3$ galaxies because we have $2$ spectra of CEERS\_01027 (Section \ref{sec:datasam}). We do not include the spectrum of CEERS\_01019, which has signatures of AGN \citep{Larson2023,Isobe2023b}, in the composite spectra at $\meanz=8$ because the $\Lya$ line width could be broadened by AGN. We note that we do not include CEERS\_01019 in the analysis of $\Lya$ line profile but include it in the analysis of $\Lya$ EW, fraction, and $\xHI$. 
\begin{figure}[t]
    \centering
    \includegraphics[scale=0.5]{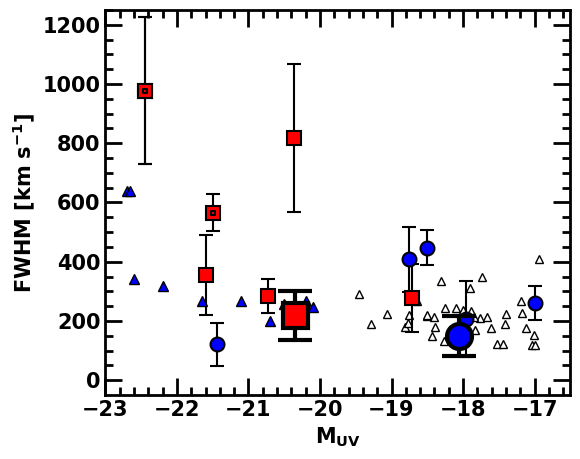}
    \caption{$\Lya$ line width as a function of absolute UV magnitude. The large blue circles and large red squares represent the FHWMs measured from the composite spectra at $\meanz=7$ and $8$, respectively. The small blue circles and small red squares denote the FHWMs of individual galaxies in our sample at $z\sim7$ and $8$, respectively. The red double squares show the FWHMs of CEERS\_01019 and GN-z11 \citep{Bunker2023a} in our sample, both of which have signatures of AGN \citep{Larson2023,Isobe2023b,Bunker2023a}. The blue and white triangles indicate the FWHMs of the galaxies at $z\sim7$ \citep{Ouchi2010,Endsley2022} and $2-3$ \citep{Hashimoto2017,Kerutt2022}, respectively.   }
    \label{fig:M_UV-FWHM}
\end{figure}
For stacking individual spectra at $z\sim7$ and $8$, we shift the galaxy spectra into the mean redshifts $\meanz=7$ and $8$ based on the systemic redshift, respectively. We then obtain the composite spectra, taking averages of the fluxes normalized with the continuum fluxes of individual spectra. We calculate $1\sigma$ errors of the composite spectra by taking averages of the error spectra of individual galaxies. The composite spectra at $\meanz=7$ and $8$ are shown in Figure \ref{fig:composite}. We conduct spectral fittings to the composite spectra in the same manner as Section \ref{subsec:fitting}. We calculate the intrinsic FWHMs of $\Lya$ emission lines from the individual spectra and composite spectra by $\mathrm{FWHM}=2\sqrt{2\log2}\sigma c/\lambda_{\Lya}(1+\mathrm{z_{sys}})$. Note that we have already corrected for instrumental broadening of line profiles by conducting convolution with the LSF of JWST/NIRSpec in the fittings (Section \ref{subsec:fitting}). The FWHM values of composite spectra at $\meanz=7$ and $8$ are $\mathrm{FWHM}=149\pm68$ km $\mathrm{s^{-1}}$ and $2   19\pm83$ km $\mathrm{s^{-1}}$, respectively. In Figure \ref{fig:M_UV-FWHM}, we compare FWHMs measured in this study with those taken from the literature. The FWHMs of galaxies at $z\sim7-8$ in this work are comparable with those at $z\sim7$ \citep{Ouchi2010,Endsley2022} and lower redshift $z\sim2-3$ \citep{Hashimoto2017,Kerutt2022}. We find no large evolution of FWHMs from $z\sim2$ to $8$, which is consistent with that between $z=5.7$ and $6.6$ \citep{Ouchi2010}. In Figure \ref{fig:profile_evolution}, we compare the composite spectra at $\meanz=7$ with those at $\meanz=8$. While we find no large evolution of $\Lya$ line profiles, we find a evolution of the emission line flux. Because the composite spectra are normalized with their continuum fluxes, the evolution of the emission line flux suggests the evolution of the $\Lya$ EW.

 \begin{figure}[t]
    \centering
    \includegraphics[scale=0.55]{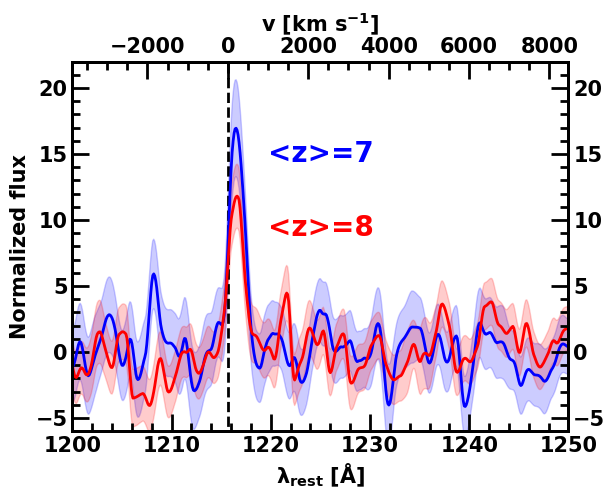}
    \caption{Evolution of $\Lya$ line profiles. The top and bottom axes are the velocity offset and rest-frame wavelength. The y-axis corresponds to the fluxes normalized with the continuum fluxes. The blue and red solid lines represent the composite spectra of the $\meanz=7$ and $8$ galaxies, respectively. The light shade regions show the $1\sigma$ errors of the composite spectra. The vertical black dashed line denotes the rest-frame $\Lya$ wavelength.}
    \label{fig:profile_evolution}
\end{figure}

\subsection{Ly$\alpha$ Equivalent Width}
\label{subsec:LyaEW}

We obtain $\Lya$ fluxes $F_{\Lya}$ by the equation,
\begin{align}
 F_{\Lya}=\int_{\lambda_\mathrm{min}}^{\lambda_\mathrm{max}}d\lambda\qty[Ae^{-\qty(\frac{\lambda-\lambda_\mathrm{cen}}{\sigma})^2}-a\lambda^{-\beta}]\times e^{-\tau(\lambda_\mathrm{obs})}
\end{align}
where $\lambda_\mathrm{max}$ ($\lambda_\mathrm{min}$) is the maximum (minimum) value of the observed wavelength for the $\Lya$ emission. We estimate flux errors from the MCMC results on the basis of the error spectrum made in \citet{Nakajima2023}, \citet{Harikane2024}, and \citet{Bunker2023b}. The rest-frame Ly$\alpha$ equivalent width $\EWLya$ is calculated with the equation, 
\begin{align}
\EWLya=\frac{F_\mathrm{Ly\alpha}}{f_c(1+z_\mathrm{sys})}
\label{eq.3}
\end{align}
where $f_c$ is the continuum fluxes at the observed wavelengths, $f_c=a\lambda_\mathrm{cen}^{-\beta}$. 
For the galaxies with no $F_\mathrm{Ly\alpha}$ measurements (i.e., no Ly$\alpha$ detections), we estimate upper limits of the $\Lya$ fluxes. For each galaxy, we conduct Monte-Carlo simulations with 100 mock spectra that are random realizations of the observed spectrum with the error spectrum. We conduct the model fitting to the 100 mock spectra to obtain Ly$\alpha$ flux measurements in the same manner as Section \ref{subsec:fitting}, and determine the $1\sigma$ error of the $\Lya$ flux with the distribution of the Ly$\alpha$ flux measurements of the 100 mock spectra. From these $\Lya$ flux errors, we calculate $3\sigma$ upper limits of $\EWLya$ for the galaxies with no $\Lya$ detections. 
\begin{figure}[t]
    \centering
    \includegraphics[scale=0.55]{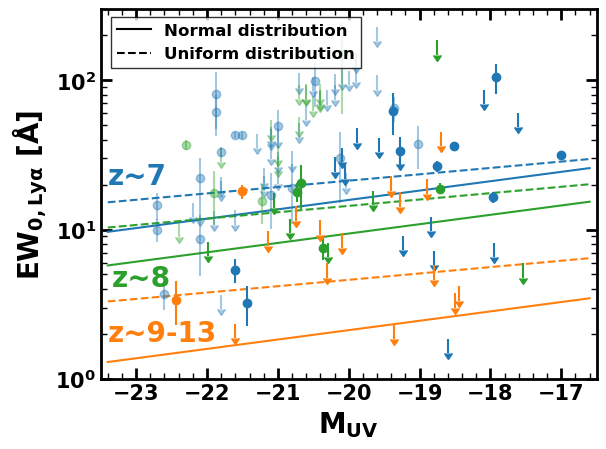}
    \includegraphics[scale=0.55]{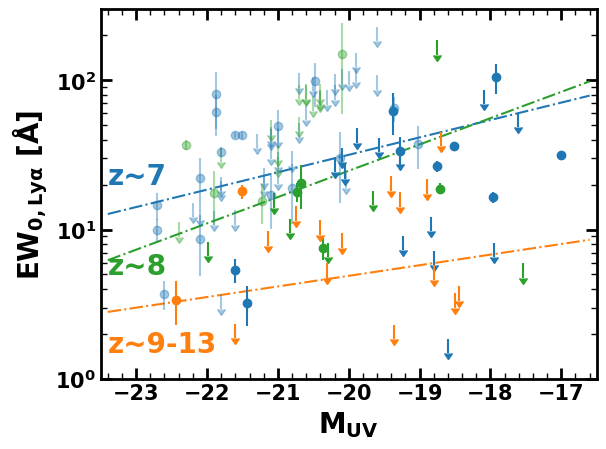}
    \caption{Rest-frame $\Lya$ equivalent width as a function of absolute UV magnitude. The blue, green, and orange circles (arrows) show the measurements ($3\sigma$ upper limits) of the galaxies at $z\sim7$, $8$, and $9-13$ of our sample, respectively. The light blue, light green, and light orange symbols are the same as the blue, green, and orange symbols, but for the galaxies obtained from the literature (\citealt{Ono2012,Schenker2012,Oesch2015,Song2016,Shibuya2018,Matthee2019,Tilvi2020,Endsley2022,Jung2022,Kerutt2022}; see Appendix C of \citealt{Jones2023}). The top and bottom panels present the results of the linear-function and \texttt{cenken} function fittings, respectively. In the top panel, the solid (dashed) lines denote the best-fit linear functions obtained by adopting a normal (uniform) distribution for the $\EWLya$ upper limits. See Section \ref{subsec:LyaEW} for details of the fittings.}
    \label{fig:M_UV-EW}
\end{figure}
We compare our measurements of $\EWLya$ with those in the literature \citep{Tang2023,Jones2023,Saxena2023,Chen2024} in Table \ref{tab:comparison} and find good agreement except for JADES\_10013682, CEERS\_80374, and CEERS\_ 80239. For JADES\_10013682, the $\EWLya$ value measured in this study ($31.5\pm10.0$ \AA) is lower than those in \citet{Jones2023} ($147.8\pm19.6$ \AA) and \citet{Saxena2023} ($337.2\pm175.5$ \AA). While the $\EWLya$ value of \citet{Jones2023} shown in Table \ref{tab:comparison} is derived by combining the $\Lya$ flux measured from the grating spectrum in \citet{Saxena2023} and the continuum flux measured from the prism spectrum in \citet{Jones2023}, our $\EWLya$ value is derived based on the grating spectrum alone. Our measurement of the $\Lya$ flux from the grating spectrum, $(2.0\pm0.2)\times10^{-18}\ \mathrm{erg}\ \mathrm{s}^{-1}\ \mathrm{cm}^{-2}$, is consistent with that reported by \citet{Saxena2023}, $(2.2\pm0.5)\times10^{-18}\ \mathrm{erg}\ \mathrm{s}^{-1}\ \mathrm{cm}^{-2}$. We thus confirm that the differences of the $\EWLya$ measurements of JADES\_10013682 come from the continuum fluxes obtained from the spectra with different resolution, the grating and prism spectra although our $\Lya$ EW measurement is explained by $2\sigma$ error of those in \citet{Saxena2023}. The impacts of the choice of the grating and prism spectra on $\Lya$ EW measurements are discussed in \ref{sec:apA}. For CEERS\_80374, we do not detect $\Lya$ emission at $>3\sigma$ significance level and measure only $3\sigma$ upper limits of $\EWLya$, $<86.7$ \AA, while \citet{Chen2024} report a $\Lya$ detection with $\EWLya=205^{+48}_{-27}$ \AA. For CEERS\_80239, our measurements ($105.3\pm72.1$ \AA) are lower than those of \citet{Chen2024} ($334^{+109}_{-62}$ \AA). These differences are due to the aperture to produce the spectra. Our spectra are produced via the summation of $3$ pixels ($0.3\arcsecond$) along the spatial direction centered on the spatial peak position to minimize the effects coming from the noisy regions close to the edge \citep{Nakajima2023}, while the spectra of \citet{Chen2024} are extracted with the aperture that is typically $6$ pixels. Because the aperture of our spectra are narrower than those of \citet{Chen2024}, we may miss the extended fluxes of $\Lya$ halos. Our measurements thus show no $\Lya$ detection or the smaller $\EWLya$ value than that of \citet{Chen2024}. In Appendix \ref{sec:apB}, we investigate the effects of the potential flux loss of $\Lya$ halo on our results. These different $\EWLya$ measurements from those in the literature for a small number of galaxies may not affect our major results.  

In Figure \ref{fig:M_UV-EW}, we plot $\EWLya$ as a function of $M_\mathrm{UV}$. To investigate the redshift evolution of $\EWLya$, we divide our sample into three subsamples by redshift so that each subsample has at least ten galaxies and a redshift range $\Delta z\gtrsim1$. Our subsamples consist of $22$, $12$, and $17$ galaxies whose redshift ranges are $6.5\leq z<7.5$, $7.5\leq z<8.5$, and $8.5\leq z<13.5$, refered to as $z\sim7$, $8$, and $9-13$, respectively. Note that we remove JADES\_00008115 and CEERS\_01163 from our subsamples because their $M_\mathrm{UV}$ values are not determined. We perform linear-function fittings to our subsamples combined with samples of galaxies taken from the literature (\citealt{Ono2012,Schenker2012,Oesch2015,Song2016,Shibuya2018,Matthee2019,Tilvi2020,Endsley2022,Jung2022,Kerutt2022}; see Appendix C of \citealt{Jones2023}), which do not overlap the galaxies in this study. For the $3\sigma$ upper limits of $\EWLya$, we adopt two different distributions in the fittings. First, we adopt normal distributions with a mean value of $0$ and standard deviations of the $1\sigma$ limits of $\EWLya$. Second, we adopt uniform distributions between $0$ and $3\sigma$ upper limits of $\EWLya$. We first fit to the galaxies at $z\sim7$ to determine the slope and the intercept. We then use the fixed slope at $z\sim7$ and determine the intercept for the galaxies at $z\sim8$ and $9-13$. The best-fit linear functions are presented in the top panel of Figure \ref{fig:M_UV-EW}. There is no significant difference between the results using normal distributions and uniform distributions.  
We also conduct fittings based on the Kendall correlation test, using the \texttt{cenken} function in the NADA library from the R-project statistics package built on \citet{Akritas1995} and \citet{Helsel2005}. We use the \texttt{cenken} function because it computes the Kendall rank correlation coefficient and an associated linear function for censored data, such as a set of data including upper limits. The fitting results using the \texttt{cenken} function are shown in the bottom panel of Figure \ref{fig:M_UV-EW}. In both of the top and bottom panel of Figure \ref{fig:M_UV-EW}, we find that $\EWLya$ becomes lower at higher redshift, which is a clear signature of the redshift evolution of $\xHI$ with observational data alone. The evolution of $\Lya$ EW is prominent between $z\sim8$ and $9-13$, suggesting that reionization significantly proceeded between $z\sim8$ and $9-13$. The p-values for the hypothesis that there is no correlation between $M_\mathrm{UV}$ and $\EWLya$ at $z\sim7$, $8$, and $9-13$ are $p=0.58$, $0.98$, and $0.96$, respectively. These p-values indicate there is no clear correlation between $M_\mathrm{UV}$ and $\EWLya$ at each redshift bin.

\subsection{Ly$\alpha$ Fraction}
\label{subsec:XLya}

\begin{figure}[t]
    \centering
    \includegraphics[scale=0.55]{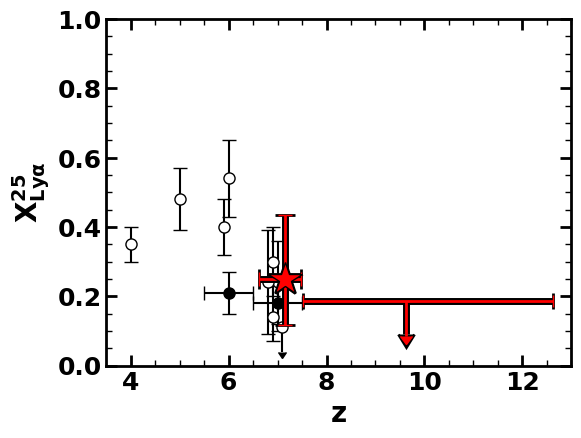}
    \includegraphics[scale=0.55]{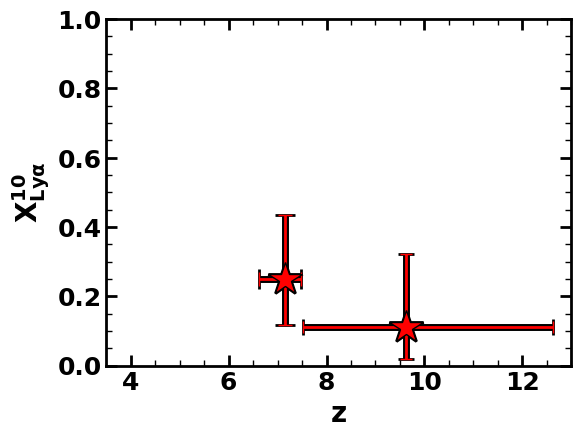}
    \caption{$\Lya$ fraction as a function of redshift. Top: the red star mark and black circles indicate the $\XLya^{25}$ values obtained from the galaxies, all of which are spectroscopically confirmed in this work and \citet{Jones2023}, respectively. The red arrow denote the same as the red star mark, but for the upper limit of $\XLya^{25}$. The white circles represent the $\XLya^{25}$ values obtained from the galaxies including those confirmed via the photometry in the literature \citep{Stark2011,Ono2012,Schenker2012,Schenker2014,Pentericci2018,Mason2019}. Bottom: the red star marks show the $\XLya^{10}$ values obtained in this work.}
    \label{fig:z-XLya}
\end{figure}

To evaluate the ionizing state of the IGM, we calculate the $\Lya$ fraction. We define the $\Lya$ fraction $\XLya^{\mathrm{EW_{th}}}$ by the ratio of the number of the galaxies with strong $\Lya$ emission ($\mathrm{EW}>\mathrm{EW_{th}}$) to the number of total galaxies. The $\Lya$ fraction $\XLya^{25}$ is typically investigated for galaxies with $-21.75<\mathrm{M_{UV}}<-20.25$ and $-20.25<\mathrm{M_{UV}}<-18.75$ (e.g., \citealt{Stark2011,Ono2012,Schenker2012,Schenker2014,Pentericci2018,Jones2023}). We have only small numbers of galaxies with $-20.25<\mathrm{M_{UV}}<-18.75$ at $z\sim8$ and $9-13$, and thus investigate the $\Lya$ fraction at rebined redshift $z\sim7$ and $8-13$. For 12 (9) galaxies  with $-20.25<\mathrm{M_{UV}}<-18.75$ at $z\sim7$ ($8-13$), we find 3 (0) galaxies with $\mathrm{EW}>25$ \AA. We obtain $\XLya^{25}=0.25_{-0.13}^{+0.19}$ (3/12) and $<0.19$ (0/9) at $z\sim7$ and $8-13$, respectively. We estimate the $1\sigma$ error and upper limit, using the statistics for small numbers of events by \citet{Gehrels1986}. As shown in top panel of Figure \ref{fig:z-XLya}, our $\XLya^{25}$ value at $z\sim7$ is consistent with those in the literature \citep{Ono2012,Schenker2012,Schenker2014,Pentericci2018,Jones2023}. We note that the denominators of $\XLya^{25}$ (i.e., the LBGs) are confirmed via the photometry in the literature, while all of the galaxies to calculate $\XLya^{25}$ is spectroscopically confirmed in this work and \citet{Jones2023}. Previous studies have indicated that $\XLya^{25}$ increases with redshift at $4\leq z\leq6$, while $\XLya^{25}$ sharply decreases between $z=6$ and $7$. Our $\XLya^{25}$ value is consistent with the trend of this sharp decrease in $\XLya^{25}$ at $6\leq z\leq7$, which is due to the poor $\Lya$ transmission in the moderately neutral IGM. At $z\sim8-13$, we only obtain upper limit of $\XLya^{25}$ due to no galaxy with $\mathrm{EW}>25$ \AA. Because the $\Lya$ EW becomes lower at higher redshift as shown in Figure \ref{fig:M_UV-EW}, we need a lower threshold of EW than $25$ \AA\ to detect $\Lya$ emitting galaxies at $z\gtrsim8$. We thus calculate the $\Lya$ fraction $\XLya^{10}$ for galaxies with $-20.25<\mathrm{M_{UV}}<-18.75$. We find one galaxy with $\mathrm{EW}>10$ \AA\ at $z\sim8-13$ and obtain $\XLya^{10}=0.25_{-0.13}^{+0.19}$ (3/12) and $<0.11_{-0.09}^{+0.21}$ (1/9) at $z\sim7$ and $8-13$, respectively. In the bottom panel of figure \ref{fig:z-XLya}, we find the trend of decrease in the $\Lya$ fraction at higher redshift between $z\sim7$ and $8-13$. In summary, combining the results of $\XLya^{25}$ in this work and literature with those of $\XLya^{10}$ in this work, we find that the $\Lya$ fraction monotonically decrease from $z\sim6$ to $8-13$ because of the more neutral hydrogen in the IGM at the earlier epoch of reionization. As for $\xHI$ estimation, the full distribution (i.e., the probability distribution function) of $\Lya$ EW allow us to make stronger constraints on $\xHI$ than $\XLya$ \citep{Treu2012,Treu2013,Mason2018} as described in Section \ref{sec:intro}. We thus mainly focus on $\xHI$ estimated with the probability distribution function (PDF) of $\Lya$ EW in Section \ref{sec:xHI}.

\section{Estimates of neutral hydrogen fraction} 
\label{sec:xHI}

\begin{figure}[t]
    \centering
    \includegraphics[scale=0.6]{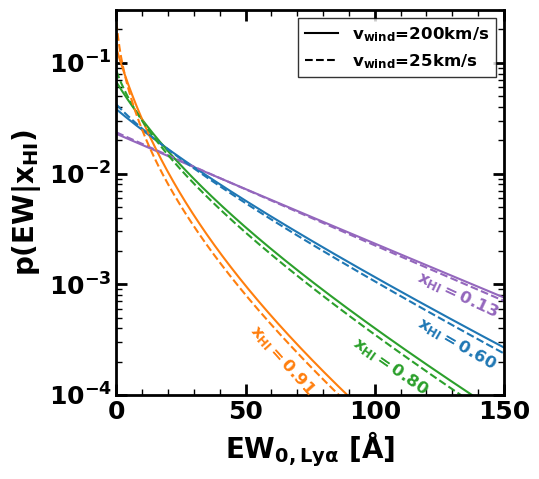}
    \includegraphics[scale=0.6]{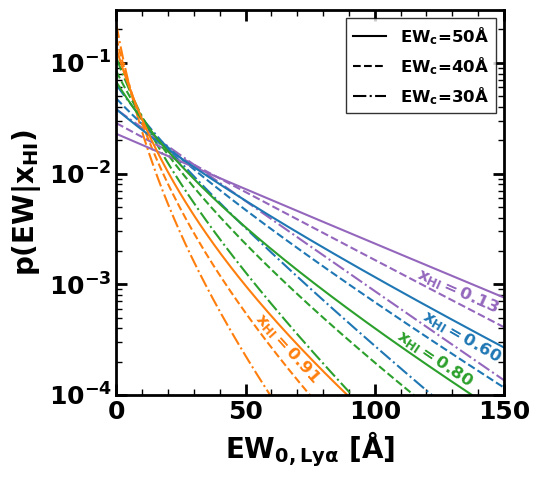}
    \caption{Probability distribution functions of the rest-frame $\Lya$ EW. The purple, blue, green, and orange solid lines indicate the EW distribution models \citep{Dijkstra2011} with a neutral hydrogen column density $N_\mathrm{\sc HI}=10^{20}$ $\mathrm{cm^{-2}}$, an outflow velocity $v_\mathrm{wind}=200$ km $\mathrm{s}^{-1}$, a scale factor $\mathrm{EW_c}$=50 \AA, and neutral hydrogen fractions $\xHI=0.13$, $0.60$, $0.80$, and $0.91$, respectively. In the top panel, the dashed lines represent the same as the solid lines, but for the models with $v_\mathrm{wind}=25$ km $\mathrm{s}^{-1}$. In the bottom panel, dashed (dashed-dotted) lines denote the same as the solid lines, but for the models with $\mathrm{EW_c}$=40 (30) \AA}.
    \label{fig:EW-PDF}
\end{figure}

We estimate neutral hydrogen fractions $\xHI$, comparing the rest-frame $\Lya$ EW values (simply referred to as EW in this section) obtained from observations with those of theoretical EW distribution models developed by \citet{Dijkstra2011}. As shown in Figure \ref{fig:EW-PDF}, EW values at low $\xHI$ universe are higher than those at high $\xHI$ universe, due to the $\Lya$ damping wing absorption made by the IGM. In the EW distribution models, \citet{Dijkstra2011} quantify the PDF of the fraction of $\Lya$ photons transmitted through the IGM $T_\mathrm{IGM}$, combining galactic outflow models with large-scale seminumeric simulations of reionization. There are two parameters, the neutral hydrogen column density $N_\mathrm{\sc HI}$ and the outflow velocity $v_\mathrm{wind}$, in the galactic outflow models. The EW distribution models are constructed under the assumption that the IGM at $z=6$ is completely transparent to $\Lya$ photons, and that the PDF of EW changes only by the evolution of the ionization state of the IGM. It is also assumed that the EW distribution at $z=6$ is described by an exponential function, $p_{z=6}(\mathrm{EW})\propto\exp(\mathrm{-EW/EW_c})$ where $\mathrm{EW_c}$ is a scale factor of EW at $z=6$. The EW distributions during the epoch of reionization are computed with $p(\mathrm{EW}|\xHI)=N\int_0^1dT_\mathrm{IGM}p(T_\mathrm{IGM})p_{z=6}(\mathrm{EW}/T_\mathrm{IGM})$ where $N$ is a normalization factor, and $p(T_\mathrm{IGM})$ is a PDF of $T_\mathrm{IGM}$ which is computed in \citet{Dijkstra2011}. As a fiducial model, \citet{Dijkstra2011} use a set of the models with $N_\mathrm{\sc HI}=10^{20}$ $\mathrm{cm^{-2}}$, $v_\mathrm{wind}=200$ km $\mathrm{s}^{-1}$, and $\mathrm{EW_c}=50$ \AA, in which $T_\mathrm{IGM}$ is computed at $z\sim8.6$. For comparison, we show the models of EW distribution with $N_\mathrm{\sc HI}=10^{20}$ $\mathrm{cm^{-2}}$, $v_\mathrm{wind}=25$ and $200$  km $\mathrm{s}^{-1}$, and $\mathrm{EW_c}=30$, $40$, and $50$ \AA\ in Figure \ref{fig:EW-PDF}. The EW distribution does not largely change for different $v_\mathrm{wind}$ values, but moderately change for different $\mathrm{EW_c}$ values. Because \citet{Dijkstra2011} find the relation of $\dvLya\sim2v_\mathrm{wind}$ in the galactic outflow model, the $\Lya$ velocity offset of the models with $v_\mathrm{wind}=200$  km $\mathrm{s}^{-1}$ is corresponding to $400$ km $\mathrm{s}^{-1}$. We note that the lower $\Lya$ velocity offsets in our sample of  $\dvLya=234\pm76$ km $\mathrm{s}^{-1}$ (mean values; Section \ref{subsec:vLya}) than that of models are unlikely to affect our estimates because $v_\mathrm{wind}$ does not largely change the EW distribution as shown in the top panel of Figure \ref{fig:EW-PDF}. Recent observational studies show column densities of neutral gas of high-z galaxies \citep{Umeda2023,Heintz2023}, which are $N_\mathrm{\sc HI}\sim10^{19}-10^{23}$ $\mathrm{cm^{-2}}$. The observed values of $N_\mathrm{\sc HI}$ are comparable to those of the models. \citet{Dijkstra2011} choose a scale factor of $\mathrm{EW_c}=50$ \AA\ because the $\Lya$ fraction of LAEs with $\mathrm{EW}>75$ \AA\ estimated with the $\mathrm{EW_c}$ value is about 0.2, which corresponds to the median value observed at $z\sim6$ \citep{Stark2010}. However, \citet{Pentericci2018} claim that a scale factor of a best-fit exponential function that match observations at $z\sim6$ is $\mathrm{EW_c}=32\pm8$ \AA. \citet{Mason2018} parameterize the $z\sim6$ EW distribution \citep{DeBarros2017,Pentericci2018} as an exponential function plus a delta function which depend on UV magnitude and obtain the scale factor of $\mathrm{EW_c}=31+12\tanh[4(M_\mathrm{UV}+20.25)]$ \AA, which is corresponding to about $30$ \AA\ for the $M_\mathrm{UV}$ values of our sample. We thus use sets of the models with $N_\mathrm{\sc HI}=10^{20}$ $\mathrm{cm^{-2}}$, $v_\mathrm{wind}=200$ km $\mathrm{s}^{-1}$, and $\mathrm{EW_c}=35\pm5$ \AA, which are comparable with the observed values.  

\begin{figure}[t]
    \centering
    \includegraphics[scale=0.48]{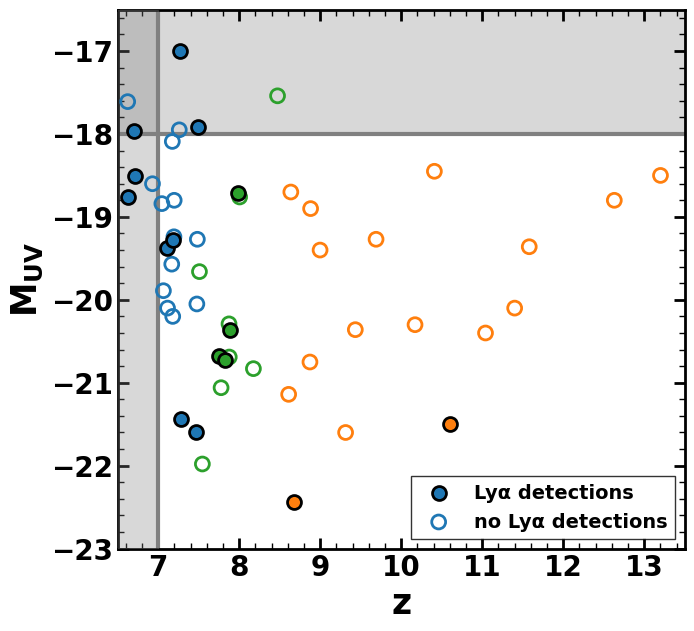}
    \caption{Absolute UV magnitude distribution of our sample galaxies as a function of redshift. The blue, green, and orange filled (open) circles denote galaxies with (no) $\Lya$ detections at $z\sim7$, $8$, and $9-13$, respectively. The grey solid lines represent the selection criteria for our sample (Section \ref{sec:xHI}).
    }
    \label{fig:z-M_UV}
\end{figure}

Figure \ref{fig:z-M_UV} presents $M_\mathrm{UV}$ distribution as a function of redshift for our sample. We find that there is a trend that faint galaxies are included for our galaxies in the low-redshift bins. To avoid systematics introduced by bias in UV magnitudes, we select galaxies, removing $8$ ($1$) galaxies at $z\sim7$ ($8$) fainter than characteristic luminosity with the criterion of $z<7$ or $M_\mathrm{UV}>-18$. We conduct two sample Kolmogorov–Smirnov tests between the UV magnitudes for the galaxies of our subsamples (Secion \ref{subsec:LyaEW}) before and after the selection. Note that the redshift range of our subsamples for $z\sim7$ after the selection is $7\leq z<7.5$. We confirm that the UV magnitude distributions are not similar between the glaxies at $z\sim7$ and $z\sim8$ before the selection at the $99\%$ significance level, but consistent for all of the 3 redshift bins after the selection at the $99\%$ significance level. We estimate $\xHI$ for our subsamples of $z\sim7$, $8$, and $9-13$, which consists of $14$ ($=22-8$), $11$ ($=12-1$), and $17$ galaxies, respectively (Section \ref{subsec:LyaEW}).

Here, we need to be careful about the effects of the slit loss on the $\Lya$ EW measurements because we compare the $\Lya$ EW obtained from JWST/NIRSpec MSA observations with the theoretical models of $\Lya$ EW distributions based on the VLT/FORS2 observations \citep{Pentericci2018} to estimate $\xHI$. Our sample galaxies are observed with NIRSpec MSA micro-shutter, which could miss the fluxes of extended $\Lya$ halos due to the small area of the micro-shutters ($0.20\arcsecond\times0.46\arcsecond$) \citep{Jakobsen2022} compared to the FORS2 slit ($\sim1\arcsecond\times8\arcsecond$) \citep{Pentericci2018}. The slit loss effects on $\Lya$ emission are discussed in recent studies. While \citet{Jung2023} claim that the $\Lya$ flux of one source measured from NIRSpec MSA is only $20\%$ of those  measured from MOSFIRE, \citet{Tang2023} find the $\Lya$ EWs of four sources measured from NIRSpec MSA are consistent with those measured from ground-based observations. \citet{Jones2023} also show that the $\Lya$ EW measurements from NIRSpec MSA are not likely to affect their results. In addition, our $\XLya^{25}$ value obtained from NIRSpec MSA observations is consistent with that obtained from FORS2 observations \citep{Pentericci2018}, suggesting that slit loss effects on $\Lya$ EW measurements may not be significant. However, we need to quantitatively asses the slit loss effects on $\Lya$ EW measurements to estimate $\xHI$ accurately. As done in \citet{Tang2024}, we conduct forward-modeling analysis to comapre $\Lya$ EW obtained from NIRSpec MSA with those from FORS2. The details of the analysis are described in Appendix \ref{sec:apB}. The results show that $\Lya$ EW measured with NIRSpec MSA is $72\pm8\%$ of those with FORS2. We thus use the corrected $\Lya$ EWs divided by this fraction ($72\pm8\%$) as fiducial values to estimate $\xHI$.

To estimate the probability distributions of $\xHI$, we perform a Bayesian inference based on that of \citet{Mason2018}. We note that our Bayesian method sets only $\xHI$ as a free parameter, instead of $\xHI$ and $M_\mathrm{UV}$ in the method of \citet{Mason2018} because our subsamples at each redshift bin have similar distributions of UV magnitude as described above. Under the Bayesian method, posterior probability distribution $p(\xHI|\mathrm{EW})$ is base on the equation, 
\begin{align}
    p(\xHI|\mathrm{EW})\propto \prod_ip(\mathrm{EW}_i|\xHI)\ p(\xHI)
\label{eq.4}
\end{align}
where $\mathrm{EW}_i$ represents EW measurements of individual galaxies. Here, $p(\mathrm{EW}_i|\xHI)$ is a likelihood function, and $p(\xHI)$ is a uniform prior with $0\leq\xHI\leq1$. The likelihood functions are calculated, including the measured 1 $\sigma$ errors $\sigma_i$ by the equation, 
\begin{align}
    p(\mathrm{EW}_i|\xHI)=\int_0^\infty d\mathrm{EW}\ \frac{e^{-\frac{(\mathrm{EW}-\mathrm{EW}_i)^2}{2\sigma_i^2}}}{\sqrt{2\pi}\sigma_i}p(\mathrm{EW}|\xHI)
\label{eq.5}
\end{align}
where $p(\mathrm{EW}|\xHI)$ presents the EW-distribution model as a function of $\xHI$. For galaxies with no $\Lya$ detections, we instead use the likelihood functions including $3\sigma$ upper limits $\mathrm{EW_{lim}}$ given by
\begin{align}
    p&(\mathrm{EW}_i<\mathrm{EW_{lim}}|\xHI)=\int_{-\infty}^{\mathrm{EW_{lim}}}d\mathrm{EW}_i\ p(\mathrm{EW}_i|\xHI)\notag\\
    =&\int_0^\infty d\mathrm{EW}\ \frac{1}{\sqrt{\pi}}\int_{-\infty}^{\mathrm{EW_{lim}}} d\frac{\mathrm{EW}_i}{\sqrt{2}\sigma_i}\ e^{-\frac{(\mathrm{EW}-\mathrm{EW}_i)^2}{2\sigma_i^2}}p(\mathrm{EW}|\xHI)\notag\\
    =&\int_0^\infty d\mathrm{EW}\ \frac{1}{2}\mathrm{erfc}\left(\frac{\mathrm{EW}-\mathrm{EW_{lim}}}{\sqrt{2}\sigma_i}\right)p(\mathrm{EW}|\xHI),
\label{eq.6}
\end{align}
where erfc($x$) is the complementary error function for $x$. 
\begin{figure}[t]
    \centering
    \includegraphics[scale=0.5]{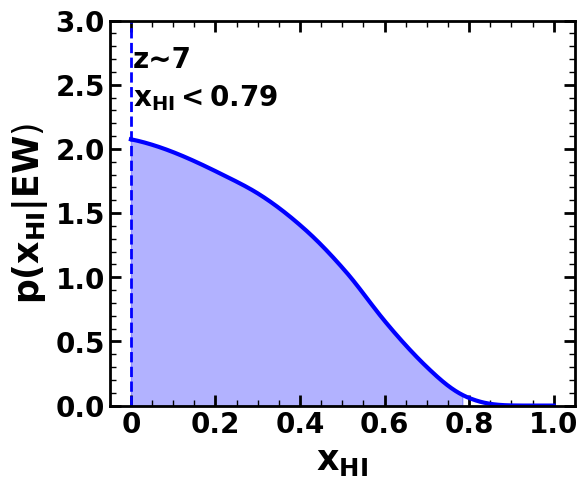}
    \includegraphics[scale=0.5]{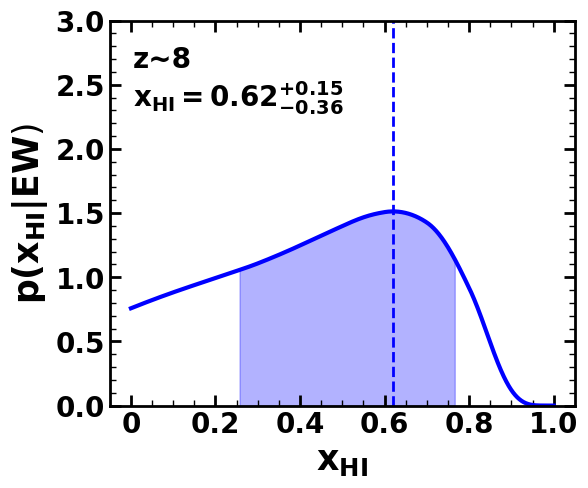}
    \includegraphics[scale=0.5]{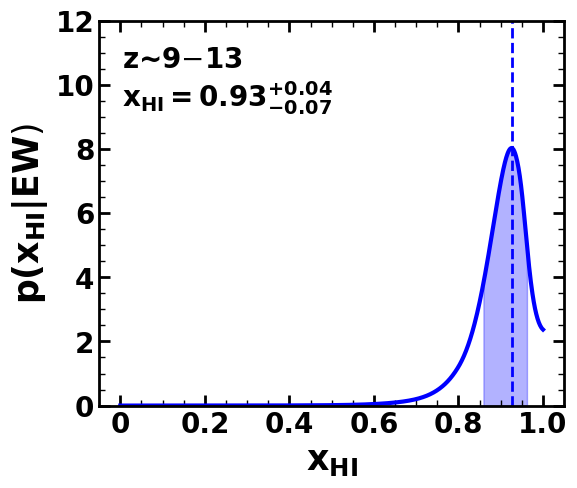}
    \caption{Posterior distributions of $x_{\mathrm{HI}}$ at $z\sim7$ (top), $8$ (middle), and $9-13$ (bottom). The dashed lines show the mode, while the shaded regions represent the $99.7$th, $68$th, and $68$th percentile highest posterior density intervals (HPDI) at $z\sim7$, $8$, and $9-13$, respectively. The estimated values are $\xHI<0.79$, $=0.62^{+0.15}_{-0.36}$, and $0.93^{+0.04}_{-0.07}$} at $z\sim7$, $8$, and $9-13$, respectively.
    \label{fig:xHI}
\end{figure}

\begin{figure}[h]
    \centering
    \includegraphics[scale=0.15]{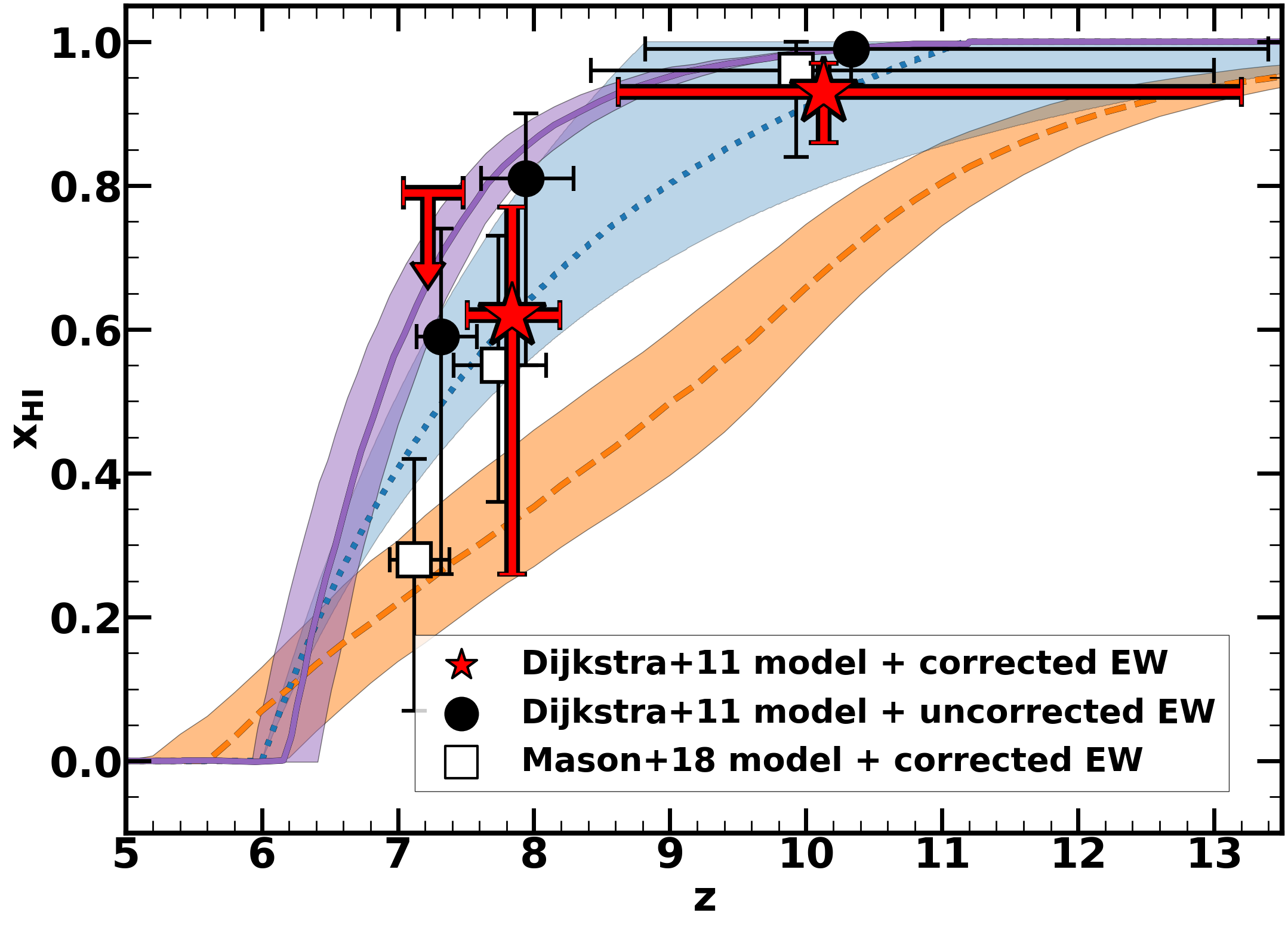}
    \caption{Comparison of the redshift evolution of $\xHI$. The red star marks, black circles, and white squares represent $\xHI$ estimates obtained with the \citet{Dijkstra2011} models+corrected $\Lya$ EW, the \citet{Dijkstra2011} models+un corrected $\Lya$ EW, and the \citet{Mason2018} models+corrected $\Lya$ EW, respectively.
    The purple solid, blue dotted, and orange dashed line represent the reionization scenarios suggested by \citet{Naidu2020}, \citet{Ishigaki2018}, and \citet{Finkelstein2019}, respectively. The light shade regions show the uncertainties of the reionization scenarios.}
    \label{fig:comparison}
\end{figure}

Combining the equation (\ref{eq.4})--(\ref{eq.6}), we calculate the posterior distributions of $\xHI$. As described above, we use the EW distribution models with $N_\mathrm{\sc HI}=10^{20}$ $\mathrm{cm^{-2}}$, $v_\mathrm{wind}=200$ km $\mathrm{s}^{-1}$, and $\mathrm{EW_c}=35\pm5$ \AA. The uncertainties for the scale factor of $\mathrm{EW_c}=35\pm5$ \AA\ in the EW distribution models are taken into account according to the following steps. We first calculate the posterior distributions for each of the $\mathrm{EW_c}$ values in $\Delta\mathrm{EW_c}=1$ \AA\ bins. We then take averages of the posterior distributions, using normal distributions with $\mathrm{EW_c}=35\pm5$ \AA. The posterior distributions of $\xHI$ at each redshift bin are shown in Figure \ref{fig:xHI}. We determine the $\xHI$ values and $1\sigma$ errors from the posterior distributions in the same manner as Section \ref{subsec:fitting}. The $\xHI$ values estimated with the corrected EWs are $\xHI<0.79$, $=0.62^{+0.15}_{-0.36}$, and $0.93^{+0.04}_{-0.07}$ at $z\sim7$, $8$, and $9-13$, respectively. We present a $3\sigma$ upper limit for $\xHI$ at $z\sim7$. For comparison, we also estimate $\xHI$ with two sets of models and measurements. One is the combination of EW distribution models of \citet{Dijkstra2011} and the uncorrected EWs. The other is the combination of EW distribution models of \citet{Mason2018} (Figure 7 of \citealt{Mason2018}) with the corrected EWs. As described above, the $z\sim6$ EW distribution in \citet{Mason2018} is based on LAEs observation with FORS2 \citep{DeBarros2017,Pentericci2018}. We thus estimate $\xHI$ by combining the corrected EWs with the models of \citet{Mason2018}. In Figure \ref{fig:comparison}, we present the results with the three different sets of models and EWs. The $\xHI$ estimates with uncorrected EWs are modestly overestimated compared to those with corrected EWs. In the case where we use corrected EWs, there are little differences between the $\xHI$ estimates with EW distribution models of \citet{Dijkstra2011} and those of \citet{Mason2018}. We thus adopt $\xHI$ estimates with EW distirbution models of \citet{Dijkstra2011} and corrected EWs as fiducial results.

\section{Discussion} 
\label{sec:diss}

\begin{figure*}[t]
    \centering
    \includegraphics[scale=0.32]{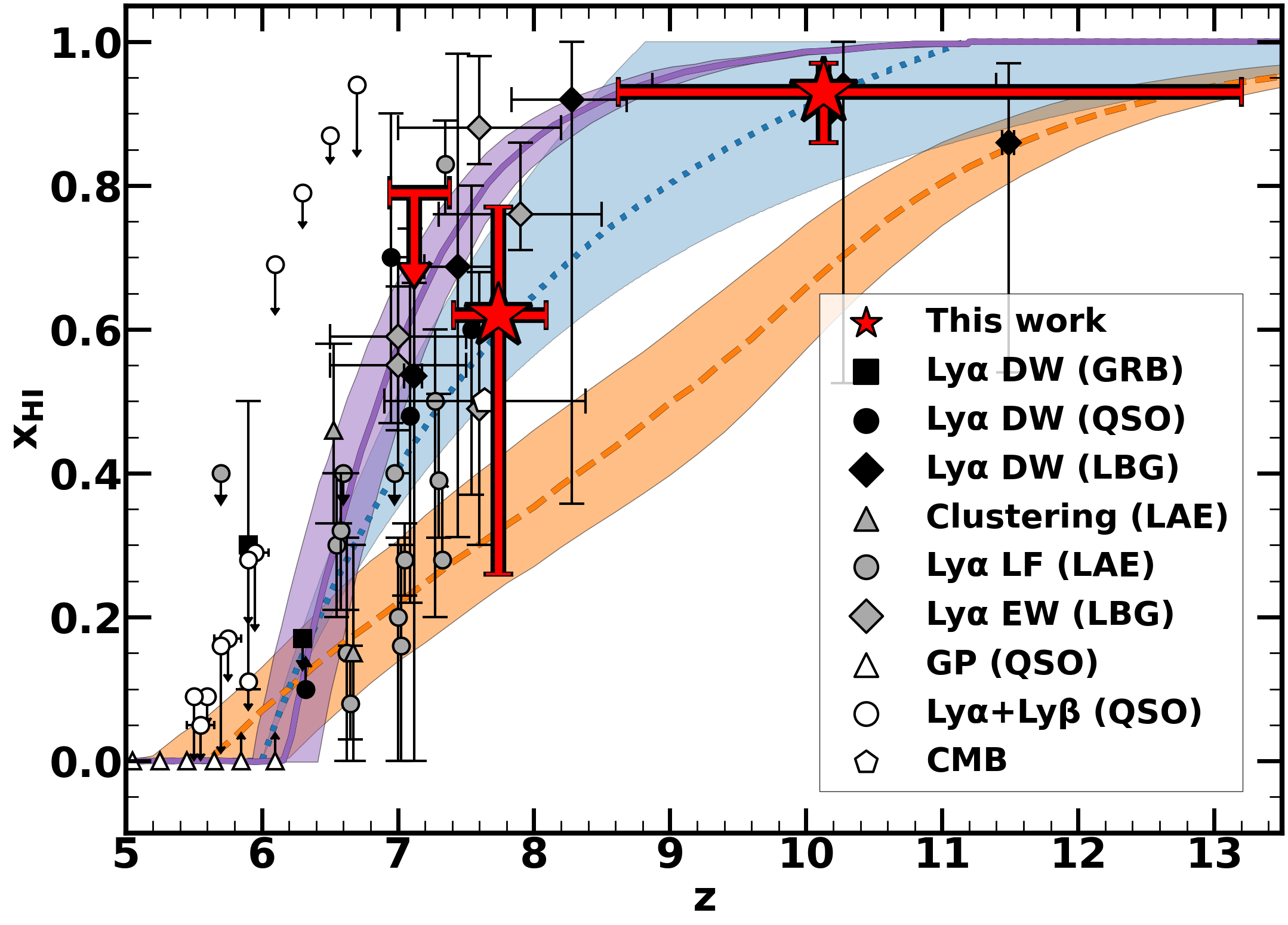}
    \caption{Redshift evolution of $\xHI$. The red star marks denote fiducial $\xHI$ estimates obtained in this work. The purple solid, blue dotted, and orange dashed line represent the reionization scenarios suggested by \citet{Naidu2020}, \citet{Ishigaki2018}, and \citet{Finkelstein2019}, respectively. The light shade regions show the uncertainties of the reionization scenarios.
    The black squares, circles, and diamonds present the $\xHI$ estimates derived from $\Lya$ damping wing absorption of GRBs (\citealt{Totani2006,Totani2014}), QSOs (\citealt{Schroeder2013}; \citealt{Davies2018}; \citealt{Greig2019}; \citealt{Wang2020}), and LBGs (\citealt{Curtis-Lake2023}; \citealt{Hsiao2023}; \citealt{Umeda2023}), respectively.
    The grey triangles and circles indicate the $\xHI$ estimates using LAE clustering analysis (\citealt{Ouchi2018}; \citealt{U23_Thesis}) and $\Lya$ luminosity function (\citealt{Ouchi2010}; \citealt{Konno2014}; \citealt{Inoue2018}; \citealt{Morales2021}; \citealt{Goto2021}; \citealt{Ning2022}; \citealt{U23_Thesis}), respectively.
    The grey diamonds denote the $\xHI$ estimates obtained from the $\Lya$ EW distribution of LBGs (\citealt{Hoag2019}; \citealt{Mason2019}; \citealt{Jung2020}; \citealt{Whitler2020}; \citealt{Bruton2023}; \citealt{Morishita2023}).
    The white triangles and circles represent the $\xHI$ estimates derived from the Gunn-Peterson through of QSOs (\citealt{Fan2006}) and $\Lya$ and $\mathrm{Ly}\beta$ forest dark fractions of QSOs (\citealt{McGreer2015,Zhu2022,Jin2023}), respectively.
    The white pentagon shows the $\xHI$ estimates obtained from the cosmic micro wave background observations under the assumption of the instantaneous reinoization (\citealt{Planck2020}).}
    \label{fig:z-xHI}
\end{figure*}

In Figure \ref{fig:z-xHI}, we present our constraints on the redshift evolution of $\xHI$ at $z=7-13$. For comparison, we also show the $\xHI$ estimates taken from the literature. Our $\xHI$ values are consistent with those estimated with $\Lya$ damping wing absorption of QSOs / Gamma Ray Bursts (GRBs), $\Lya$ luminosity function, $\Lya$ EW of LBGs, and the Thomson scattering optical depth of the CMB within the errors. While these results constrain $\xHI$ up to $z\sim 7.5$ that is limited to the redshifts of the identified QSOs and GRBs \citep{Totani2014,Wang2019,Matsuoka2023}, a recent study of \citet{Umeda2023} extends the $\xHI$ measurements to $z\sim 8-12$ with JWST galaxies via the UV continuum modeling for the $\Lya$ damping wing absorption features. The sample of \citet{Umeda2023} consists of 26 galaxies that are used in their $\Lya$ damping wing analyses, while our sample of the 53 galaxies includes all the \citeauthor{Umeda2023}'s 26 galaxies. Although there is an overlap of the samples between our and \citeauthor{Umeda2023}'s studies, our study focuses on Ly$\alpha$ line equivalent width statistics compared with the numerical simulations, which complement the study of \citet{Umeda2023} in the independent method to estimate $\xHI$. We confirm that the $\xHI$ estimates of our and \citeauthor{Umeda2023}'s studies are consistent within the errors. This consistency solidifies the validity of our measurements of $\Lya$ EW that is used to infer the $\xHI$ values. We plot the late, medium-late, and early scenarios suggested by \citet{Naidu2020}, \citet{Ishigaki2018}, and \citet{Finkelstein2019}, respectively (Section \ref{sec:intro}). Our results are consistent with the late and medium-late scenarios. 

As explained in Section \ref{sec:intro}, the late and medium-late scenarios suggest that sources of reionization form at the late epoch probably at $z\sim 7-8$. If the late scenario is correct, the slopes of UV luminosity functions should be as shallow as $\alpha>-2$ under the assumption that $f_\mathrm{esc}$ depends on the star-formation rate surface density $\Sigma_\mathrm{SFR}$ (Section \ref{sec:intro}). However, the recent observational results prefer a steep slope $\alpha<-2$ \citep{Bouwens2017,Livermore2017,Atek2018,Ishigaki2018,Finkelstein2019}. Because the requirement of the shallow UV slope does not agree with the observational results, this discrepancy would possibly suggest that the escape fraction does not depend on $\Sigma_\mathrm{SFR}$ but on the other quantities. For example, if escape fraction positively correlates with halo masses (e.g., \citealt{Gnedin2008,Wise2009,Sharma2016}), the positive correlation could lead to the late reionization with the steep UV slope suggested by the UV luminosity function observations. If we consider the possibility of star-forming galaxies as major sources of reionization, ionizing photons are prevented from escaping to the IGM because of the enrichment of metals and dusts in star-forming galaxies. To account for the high escape fractions of star-forming galaxies hosted by moderately massive halos, channels in the ISM produced by the star bursts and frequent supernovae (e.g., \citealt{Paardekooper2015,Steidel2018}) or unknown physical processes may be needed. In fact, \citet{Paardekooper2015} show that young star-forming galaxies in massive halos would have high escape fraction if supernova feedback causes either an inhomogeneous density or removal of the dense gas around the galaxies. Another possibility may be that galaxies in massive halos harbor AGNs that efficiently produce ionizing photons. While QSOs (i.e., bright AGNs) make negligible contributions to reionization due to their low number density \citep{Onoue2017,McGreer2018,Parsa2018,Wang2019,Jiang2022}, some studies suggest that the contributions from faint AGNs to reionization may be significant \citep{Haardt&Salavaterra2015,Giallongo2015,Giallongo2019}. In fact, many faint AGNs are found at $z=4-10$ by JWST observations \citep{Kocevski2023,Ubler2023,Harikane2023,Matthee2023,Maiolino2023a,Maiolino2023b}. \citet{Harikane2023} and \citet{Maiolino2023b} suggest that the contributions from faint AGNs ($M_\mathrm{UV}\gtrsim-22$) to the ionizing photons are not so large, but exist. Such many faint AGNs might cause the late reionization. In summary, our results together with the recent observational results suggest that escape fraction may positively correlates with halo masses and that the possible major reionization sources are star-forming galaxies or faint AGNs hosted by moderately massive halos. In fact, the galaxies with high neutral hydrogen column densities which disturb the escape of ionizing photons are found at $z\gtrsim8$ \citep{Heintz2023,Umeda2023,D'Eugenio2023}, indicating that significant reionization is caused by the objects which emerge at the late era of reionization ($z\sim7-8$) and are hosted by moderately massive halos.

\section{Summary} \label{sec:sum}
In this paper, we present our constraints on $\xHI$ from $\Lya$ EWs of galaxies at $z\sim7-13$. We select 54 galaxies at $6.6<z<13.2$ observed with JWST/NIRSpec in the multiple programs of ERS, DDT, GO, and GTO. The redshifts of the galaxies of our sample are spectroscopically confirmed by emission lines or spectral breaks. Because we are not able to conduct spectral fitting for one galaxy, we investigate the $\Lya$ properties of 53 galaxies. We obtain the $\Lya$ velocity offset, line width, and EW for 15 galaxies with $\Lya$ detections from the spectral fittings. For 38 galaxies with no $\Lya$ detections, we measure $3\sigma$ upper limits of $\Lya$ EW. Our major findings are summarized below:
\begin{itemize}
    \item[1.] We find no redshift evolution of the $\Lya$ velocity offset and line width from $z\sim2-3$ to $z\sim7-13$. While the composite spectra at mean redshifts of $\meanz=7$ and $8$ also show no evolution of line profiles, we find clear attenuation of line flux, suggesting the evolution of the $\Lya$ EW. To investigate the redshift dependence of the $\Lya$ EW, we conduct linear-function fittings for galaxies at $z\sim7$, $8$, and $9-13$ by adopting various probability distributions for the upper limits of $\Lya$ EW. We find that $\EWLya$ becomes lower at higher redshift (Figure \ref{fig:M_UV-EW}), which is a clear signature of the redshift evolution of $\xHI$ based on the observational data alone. The $\Lya$ fraction measurements at $z\sim7$ also suggest the evolution of $\xHI$ between $z\sim6$ and $7$.  

    \item[2.] To estimate $\xHI$ independent of UV magnitude, we select 42 out of 53 galaxies with $z\geq7$ and $M_\mathrm{UV}\leq-18$ that have similar distributions of $M_\mathrm{UV}$ for each redshift bin. We estimate $\xHI$ by comparing our $\EWLya$ values with those of the EW distribution models \citep{Dijkstra2011} based on a Bayesian inference \citep{Mason2018}. We correct our $\EWLya$ values by forward modeling of the slit loss effects for comparison with ground-based results. The estimates are $\xHI<0.79$, $=0.62^{+0.15}_{-0.36}$, and $0.93^{+0.04}_{-0.07}$ at $z\sim7$, $8$, and $9-13$, respectively. Our $\xHI$ values are consistent with those estimated with the $\Lya$ damping wing absorption of QSOs/GRBs, $\Lya$ luminosity function, $\Lya$ EW of LBGs, and the Thomson scattering optical depth of the CMB within errors. Our $\xHI$ values are also consistent with those estimated in a complementary approach with the UV continua of bright galaxies for damping absorption estimates \citep{Umeda2023}. The redshift evolution of our $\xHI$ values is consistent with the late or medium-late reionization scenarios. Such a late reionization history suggests that major reionization sources are the objects hosted by moderately massive halos that are in contrast to the widely accepted picture of abundant faint low-mass objects for reionization source. 
\end{itemize}


\section*{Acknowledgements} We thank Mark Dijkstra for providing us with the machine-readable table of their simulation results. We thank Roland Bacon, Yoshinobu Fudamoto, Takuya Hashimoto, Kohei Inayoshi, Akio Inoue, Daichi Kashino, Satoshi Kikuta, Takanobu Kirihara, Haruka Kusakabe, Pascal Oesch, Kazuyuki Omukai, and Hidenobu Yajima for the valuable discussions on this work. This work is based on observations made with the NASA/ESA/CSA James Webb Space Telescope. The data were obtained from the Mikulski Archive for Space Telescopes at the Space Telescope Science Institute, which is operated by the Association of Universities for Research in Astronomy, Inc., under NASA contract NAS 5-03127 for JWST. The specific observations analyzed can be accessed via \dataset[10.17909/8tdj-8n28]{https://doi.org/10.17909/8tdj-8n28}. These observations are associated with programs ERS-1324 (GLASS), ERS-1345 (CEERS), GO-1433, DDT-2750, and GTO-1210 (JADES). The author acknowledge the GLASS, CEERS, GO-1433, and DDT-2750 teams led by Tommaso Treu, Steven L. Finkelstein, Dan Coe, and Pablo Arrabal Haro, respectively, for developing their observing programs with a zero-exclusive-access period. We thank JADES team for publicly releasing reduced spectra and catalog from the JADES survey. This publication is based upon work supported by the World Premier International Research Center Initiative (WPI Initiative), MEXT, Japan, and KAKENHI (20H00180, 21H04467) through Japan Society for the Promotion of Science. This work was supported by the joint research program of the Institute for Cosmic Ray Research (ICRR), University of Tokyo.

%


\software{NumPy \citep{Harris2020}, matplotlib \citep{Hunter2007}, SciPy \citep{SciPy-NMeth2020}, Astropy \citep{Astropy2013,Astropy2018,Astropy2022}}, and emcee \citep{Foreman2013}.





\bibliographystyle{aasjournal}
\bibliography{library.bib}



\clearpage
\appendix
\restartappendixnumbering
\section{Comparison of Lyman alpha equivalent width measured with the grating and prism spectra}
\label{sec:apA}
\begin{figure}[h]
    \centering
    \includegraphics[scale=0.6]{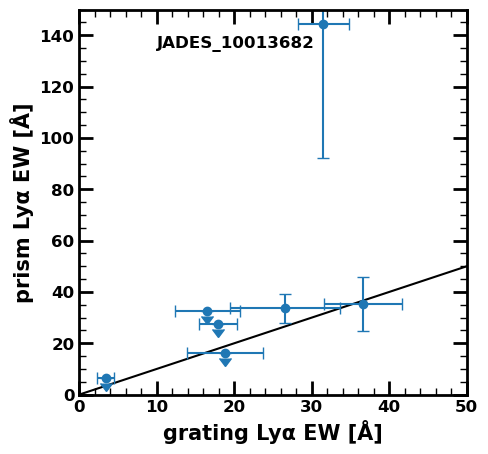}
    \includegraphics[scale=0.592]{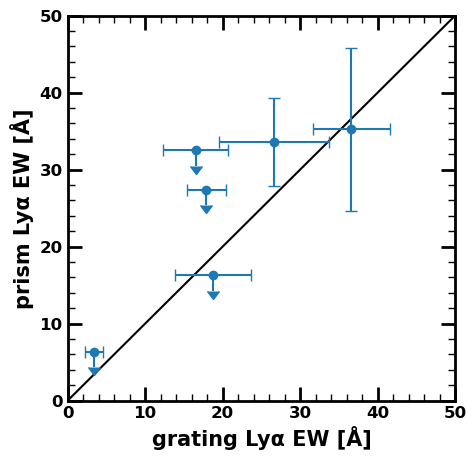}
    \caption{Comparison of $\Lya$ EW measurements with our grating spectra and those with our prism spectra. The left panel presents wide range of the figure while right panel zooms in to show the plots clearer. The circles denote $\Lya$ EWs for sources observed with both the grating and prism. The errorbars and arrows are represent the $1\sigma$ errors and $3\sigma$ upper limits, repsectively. The black line indicate that $\Lya$ EWs obtained from grating spectra are equivalent with those obtained from prism spectra.}
    \label{fig:grating-prism}
\end{figure}

In this study, we measure $\Lya$ EW with the grating and prism spectra. Recent studies of LAEs observed with JWST show that $\Lya$ emission is not detected with prism spectra, but detected with grating spectra (e.g., \citealt{Bunker2023a}; \citealt{Saxena2023}). However, $\Lya$ EWs obtained from our grating spectra are consistent with $\Lya$ EWs or $3\sigma$ upper limits of $\Lya$ EW obtained from our prism spectra except one source, JADES\_$10013682$, as shown in Figure \ref{fig:grating-prism}. Although the $\Lya$ EW measurements of JADES\_$10013682$ with the prism spectra are higher than those with the grating spectra, which is explained by $2\sigma$ errors.  Including $\Lya$ EW measurements with the prism spectra in our analysis thus would not affect our results significantly.

\section{Forward modeling}
\label{sec:apB}
\begin{figure}[h]
    \centering
    \includegraphics[scale=0.17]{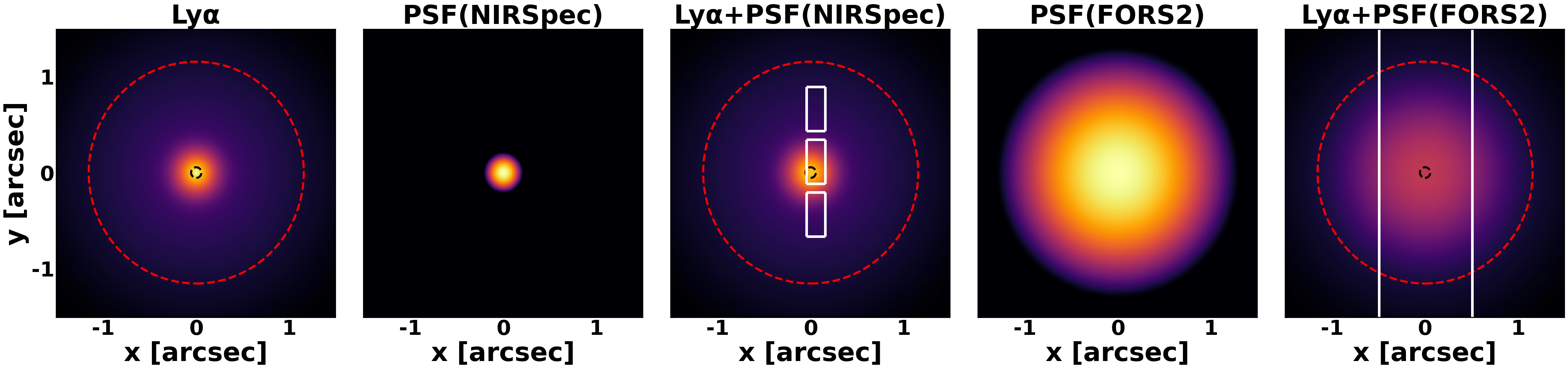}
    \includegraphics[scale=0.17]{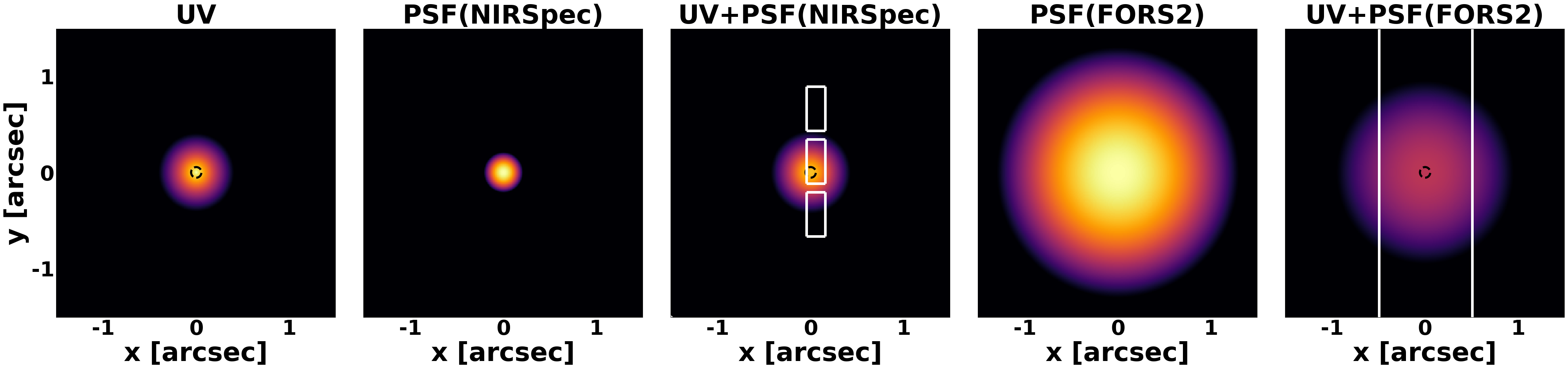}
    \caption{Example of the forward modeling of a source at $z\sim5$. From left to right in the top row, each panel indicate the intrinsic $\Lya$ spatial profile, PSF of NIRSpec, $\Lya$ emission convolved with PSF of NIRSpec, PSF of FORS2, and $\Lya$ emission convolved with PSF of FORS2. The bottom row shows the same as the top line but for UV emission. The radius of the black dashed and red dashed circles correspond to the scale lengths of the core and halo components, respectively. The white lines represent the NIRSpec MSA 3-shutter slits (middle panel) and FORS2 slits (rightmost panel).}
    \label{fig:forward_model}
\end{figure}

\begin{figure}[h]
    \centering
    \includegraphics[scale=0.6]{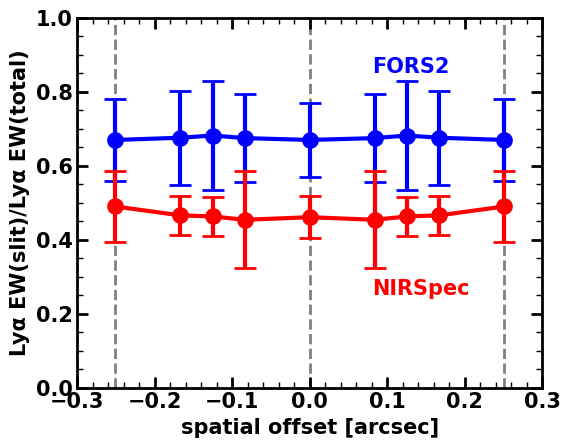}
        \includegraphics[scale=0.6]{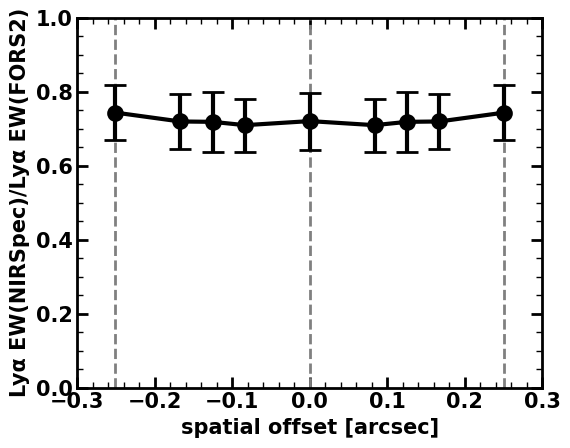}
    \caption{The fraction of $\Lya$ EW as a function of the source position. Left: the red (blue) circles represent the fractions of $\Lya$ EW measured within the NIRSpec MSA (FORS2) slit to $\Lya$ EW measured with total $\Lya$ and UV fluxes. The left, middle, and right gray dashed lines indicate the corner, center, and corner of the slit, respectively. Right: the black circles denote the fractions of $\Lya$ EW measurements with NIRSpec to those with FORS2. The gray dashed lines show the same as the top panel.}
    \label{fig:Lya_EW_fraction}
\end{figure}

As described in Section \ref{sec:xHI}, we conduct forward-modeling analysis to estimate the slit loss effects on $\Lya$ EW measurements. We first obtain the intrinsic UV and $\Lya$ spatial profile of 10 galaxies at $z\sim5-6$ based on the measurements with MUSE \citep{Leclercq2017}. For UV continuum emission, we construct a 2D exponential profile with a scale length of a core $rs_\mathrm{cont}$ and a total UV flux obtained form UV magnitude. For $\Lya$ emission, we construct a 2D exponential profile for two components, which are a core component and a halo component. The profile of the core (halo) component are obtained from a scale length of the core $rs_\mathrm{cont}$ (halo $rs_\mathrm{halo}$) and a total core (halo) $\Lya$ flux $F_\mathrm{cont}$ ($F_\mathrm{halo}$). We then convolve the UV and $\Lya$ profiles with the point spread function (PSF) of NIRSpec and FORS2. We assume the PSF is approximated by a 2D Gaussian profile. The PSF of NIRSpec is obtained from the \texttt{WebbPSF} package \citep{Perrin2014}. We adopt a seeing size $\sim0.8\arcsecond$ \citep{Pentericci2018} as the FWHM of the PSF of FORS2. In Figure \ref{fig:forward_model}, we present an example of the forward modeling for a source at $z\sim5$.
For NIRSpec, we measure $\Lya$ and UV fluxes by summing the fluxes within the aperture size $0.3\arcsecond$ (Section\ref{subsec:LyaEW}) along the spatial direction centered on the spatial peak position of the slit ($0.20\arcsecond\times0.46\arcsecond$; \citealt{Jakobsen2022}). For FORS2, we measure the fluxes by summing the fluxes within the slit ($1\arcsecond\times8\arcsecond$; \citealt{Pentericci2018}). We vary the offset of the source from the center of the slit along the diagonal line of NIRSpec MSA slit for both NIRSpec MSA and FORS2 slit. From the flux measurements with NIRSpec and FORS2, we obtain $\Lya$ EW of 10 galaxies at $z\sim5-6$. We calculate $1\sigma$ $\Lya$ EW errors by conducting Monte-Carlo simulations of 100 forward modelings with the errors of $rs_\mathrm{cont}$, $rs_\mathrm{halo}$, $F_\mathrm{cont}$, and $F_\mathrm{halo}$. We present the fractions of $\Lya$ EW measured within NIRSpec MSA (FORS2) slit to the $\Lya$ EW measured with the total $\Lya$ and UV fluxes, and fractions of $\Lya$ EW measurements with NIRSpec to those with FORS2 for different source positions in the left and right panel of \ref{fig:Lya_EW_fraction}, respectively. We confirm that the fractions do not depend on source position in the slit. We thus adopt the fraction of $\Lya$ EW measurements with NIRSpec to those with FORS2 where the source is center in the slit, $72\pm8\%$, as a correction factor in Section \ref{sec:xHI}.

\end{document}